\documentclass[traditabstract]{aa} 
\usepackage{booktabs} 
\usepackage{graphicx}
\usepackage{lmodern}
\usepackage{natbib}
\usepackage{savesym}
\usepackage{subfigure}
\usepackage{amsmath}
\savesymbol{iint}
\usepackage{txfonts}
\restoresymbol{TXF}{iint}\usepackage[colorlinks=true,     linkcolor=blue,     anchorcolor=blue,
citecolor=blue,            filecolor=blue,            menucolor=blue,
urlcolor=blue]{hyperref} 

\usepackage{xspace}

\bibpunct{(}{)}{;}{a}{}{,} 
\defcitealias{2005MNRAS.363....2K}{K05}
\defcitealias{2005A&A...441...55S}{SC05}
\defcitealias{2009A&A...506..647T}{T09}
\newcommand\simgt{\lower.5ex\hbox{$\;\buildrel>\over\sim\;$}}
\newcommand\simlt{\lower.5ex\hbox{$\;\buildrel<\over\sim\;$}}

\newcommand{\Msun}{\ensuremath{\mathrm{M}_\odot}}
\newcommand{\diff}{\ensuremath{\mathrm{d}}}
\newcommand{\deriv}[2]{\ensuremath{\frac{\diff {#1}}{\diff {#2}}}}
\newcommand{\eps}{\ensuremath{\varepsilon}}
\newcommand{\rt}{\ensuremath{\rho_\text{T}}\xspace}
\newcommand{\nth}{\textsuperscript{th}\xspace}

\newcommand {\otoprule }{\midrule[\heavyrulewidth]}
\begin{document}

\title{Mass assembly of galaxies:}

\subtitle{Smooth accretion versus mergers}
\author{B.~L'Huillier\inst{\ref{inst1},\ref{inst2}}
        \and
       F.~Combes\inst{\ref{inst1}}
        \and
        B.~Semelin\inst{\ref{inst1}}
}

\offprints{B. L'Huillier}

\institute{  
  LERMA, Observatoire de Paris, UPMC, CNRS, 61 Av.  de l'Observatoire,
  75014 Paris, France\\ \email{lhuillier@kias.re.kr,francoise.combes@obspm.fr,benoint.semelin@obspm.fr}\label{inst1}
\and
School of  Physics,  Korea  Institute for  Advanced
  Study, Seoul 130-722, Republic of Korea\label{inst2}
} 

\date{Received / Accepted 08 June 2012}

\abstract{%
Galaxies accrete their mass by means of both smooth accretion from the cosmic
web, and the mergers of smaller entities.  
We wish to quantify the respective role of these two modes of
accretion,  which could  determine the  morphological types  of galaxies
observed today. 
Multi-zoom cosmological simulations are used to estimate as a function
of time  the evolution of  mass in bound  systems, for dark  matter as
well as baryons.  
The baryonic contents of dark matter haloes are studied.
Merger histories are  followed as a function of  external density, and
the different ways in which mass is assembled in galaxies and the stellar
component accumulated are quantified.  
We  find  that  most  galaxies  assemble  their  mass  through  smooth
accretion, and only the most massive galaxies also grow significantly
through mergers. 
The  mean fraction  of mass  assembled by  accretion is  77\%,  and by
mergers 23\%. 
We present typical accretion histories of hundreds of galaxies: masses
of the most massive galaxies increase monotonically in time, mainly
through  accretion,  many  intermediate-mass objects  also  experience
mass-loss events such as tidal stripping and evaporation.  
However, our simulations suffer from the overcooling
  of massive  galaxies caused by the neglect of active galaxy
  nuclei (AGN)   feedback.
The time by  which half of the galay mass has  assembled, both in dark
matter and baryons, is a decreasing function of mass, which is compatible with
the observations of a so-called downsizing.  
At  every  epoch in  the  universe,  there  are low-mass  galaxies
actively forming  stars, while more massive galaxies  form their stars
over a shorter period of time within half the age of the universe. 
}


\maketitle

\section{Introduction} 

In the standard $\Lambda$CDM scenario, the first structures to form in
the  universe are low-mass  dark haloes,  that progressively  merge to
produce 
larger and more massive structures \citep[e.g.][]{1984Natur.311..517B}.
Baryons  then  infall into  the dark  potential  wells, forming
stars and rotationally supported galaxies \citep{1980MNRAS.193..189F}. 
Merging is  considered one of the  main mechanisms for  assembling mass in
galaxies and triggering the formation of new stars
\citep{1993MNRAS.264..201K,1998ApJ...498..504B}. 

\smallskip

Dark matter haloes and galaxies grow by both mergers and the accretion of
diffuse gas component (smooth accretion).
In recent years, advances in computational power have allowed us to study
in detail  the growth of dark  matter haloes, which can  be probed by
numerical       simulations       \citep[e.g.,][]{2008ApJ...679.1260M,
  2010MNRAS.401.2245F,    2010ApJ...719..229G,    2011MNRAS.tmp.1293T,
  2011MNRAS.413.1373W},  but  the  role  of smooth  accretion  remains
uncertain.  
Galaxy  mass  assembly  has  also  been  investigated  using  $N$-body
simulations including hydrodynamics 
\citep[e.g.,][]{2002ApJ...571....1M,2005A&A...441...55S}            and
semi-analytical models \citep[SAMs, e.g., ][]{1991ApJ...379...52W,
  1999MNRAS.305....1S,    1999MNRAS.303..188K,    2000MNRAS.319..168C,
  2003MNRAS.343...75H,    2003MNRAS.338..913H,    2006MNRAS.370..645B,
  2006MNRAS.372..933N, 2007MNRAS.375....2D,2011A&A...533A...5C}.  
\citet{2003MNRAS.345..349B} questioned the  necessity of shock heating
and found a halo mass threshold of $\simeq 10^{11} \Msun$ 
below  which the pressure  of the  shock-heated gas  is insufficiently
high to bear its own gravity and the pressure of infalling material, and is
thus unstable.  
High    resolution    simulations    based   on    either    particles
\citep{2005MNRAS.363....2K, 2009MNRAS.395..160K, 2009ApJ...694..396B, 
  2011MNRAS.417.2982F, 2011MNRAS.tmp..554V} 
or a grid~\citep{2008MNRAS.390.1326O, 2009Natur.457..451D}, have 
emphasised the coexistence of two modes of gas accretion. 
They have demonstrated that hot accretion is spherical, isotropic, and dominates
at  low redshift  and  for massive  systems,  while the  cold mode  is
anisotropic, coming from filaments, and is most significant for lower mass
galaxies and at high redshift. 
Gas accretion by means of cold streams leads to the formation of
clumps in  the disc  that fall  towards the centre  of the  galaxy and
merge     to    form     a     spheroid    \citep{2009Natur.457..451D,
  2009MNRAS.397L..64A, 2010MNRAS.404.2151C}.  
 High-redshift ($z\simeq 2$) star-forming galaxies (SFGs) have been
observed         by         integral        field         spectroscopy
\citep{2006Natur.442..786G,2009ApJ...706.1364F}.  
They seem to contain discs, which is incompatible with major mergers, but
are very clumpy.
This  may  indicate  that  they  have  a high  gas  fraction,  and  is
consistent with the 
theoretical predictions of cold accretion \citep{2009Natur.457..451D}. 
The question now  arises of the relative  roles of merging and
external accretion in galaxy  mass assembly and formation, and whether
some  combinations   of  these  processes  can   explain  the  observed
anti-hierarchical evolution of galaxies, 
which     is     usually     called     the     downsizing     process
\citep[e.g.][]{1996AJ....112..839C}.  
Through the analysis of cosmological $N$-body and hydro simulations, the
main  focus of  the present  paper is  to quantify  the  importance of
``smooth accretion'' relative to merger rates in the mass assembly of
galaxies. 

\smallskip

\label{simu}
\begin{table*}[ht!]
  \begin{center}
    \begin{tabular}{lcccc}
      \toprule
      Zoom level & 0 & 1& 2& 3 \\
      \otoprule
      $m_{\text{DM}}$  (\Msun)  &  $7.27\times  10^{10}$  &$9.09\times
      10^{9}$& $1.14\times 10^{9}$&$1.42 \times 10^{8}$ \\ 
      $m_{\text{b}}$  (\Msun)  &  $1.54\times 10^{10}$  &$1.93  \times
      10^{9}$&$2.41 \times 10^{8}$& $3.01 \times 10^{7}$\\ 
      $L_\text{box}$ (Mpc) & 137.0&68.49&34.25& 17.12\\
      $\eps_\text{soft}$ (kpc) & 50.0 & 25.0 & 12.5 & 6.25 \\
      $\diff t$ (Myr)& 20 & 10  & 5 & 2.5 \\
      $\Delta t$ (Gyr)& 0.2 & 0.2  & 0.2 & 0.1 \\
      $z_\text{end}$ & 0 &0 &0&0.46\\
      \bottomrule
    \end{tabular}
    \caption{Parameters of the multi-zoom simulation used here for the
      four levels of zoom. $L_\text{box}$  is the cube length for level 0
      zoom, and the diameter for higher level zoom; $\eps_\text{soft}$
      is the (comoving) softening parameter; $\diff t$ is the timestep; and
      $\Delta t$ is the separation between two consecutive outputs.
      \label{tab:sim_zoom}}
  \end{center}
\end{table*}

Both $N$-body simulations and SAMs have been used to
describe the  hierarchical process, by building  merging trees tracing
the formation of a given structure. 
The        Extended       Press-Schechter        (EPS)       formalism
\citep{1991ApJ...379..440B,1993MNRAS.262..627L} has  been a remarkably
successful  approximation  to obtain  mass  distributions and  merging
histories.  
However, compared to  the results derived from $N$-body simulations, there are
fundamental differences, owing to the simple hypotheses of a collapse
  independent of either the environment, the total mass of the
structure, or  the shape, although a significant  improvement was made
by             considering             ellipsoidal            collapse
\citep{2001MNRAS.323....1S,2009MNRAS.397..299M}.  
The simplifying assumptions consider the hierarchical collapse to be a
Markov  process  independent of  the  history  of  merging, while  the
reality is that it is not. 
Tracing  merging trees  from cosmological  simulations,  although more
realistic, is not a trivial task either.  
Many new algorithms have been developed and published, that have 
complementary   strengths   and   weaknesses  (the   Friend-of-Friends
algorithm (FOF),
\citealt{1985ApJ...292..371D},
SUBFIND,
\citealt{2001MNRAS.328..726S},
AdaptaHOP,
\citealt{2004MNRAS.352..376A};
\citealt{2009A&A...506..647T}, hereafter T09), and 
each relies on their own definitions and conventions to either avoid
anomalies, such as the blending of haloes and the unrealistic resulting
histories,  or  deal  with  substructures  \citep{2003MNRAS.345.1200S,
  2010MNRAS.404..502G}.  
In this paper, we use the AdaptaHOP algorithm to detect dark 
matter haloes and baryonic galaxies. 

\smallskip

Cosmological simulations show strikingly that bound structures are not
the main component of the  large-scale morphology of the universe, but
that the filamentary aspect is instead essential and the smooth component of
filaments could contain a large fraction of the mass (both dark matter (DM)
and baryons). 
Several  methods have tried to quantify  the mass in the
various components \citep{2005A&A...434..423S,2010A&A...510A..38S}, or
to    identify   the   structure    of   the    filamentary   skeleton
\citep{2006MNRAS.366.1201N,2009MNRAS.393..457S,2010MNRAS.404L..89A}. It
is essential  to estimate  the smooth accretion  mass fraction  on any
scale in $N$-body 
simulations, to more clearly understand the relative role of mergers and
accretion in galaxy formation.  
The  problem is tightly related to the
amount  of substructures  and  their evolution,  and requires  precise
definitions \citep[e.g.][]{2010MNRAS.404..502G}.

\smallskip

When  the initial density  fluctuations are  expressed by  a power-law
spectrum $P(k)  \propto k^n$, the  variance of the power spectrum on
mass  scales $M$  is proportional  to $M^{\frac{-(n+3)}{3}}$;   the
low-mass structures should then collapse first, when $n > -3$. 
Since this  is the  case in our  universe, hierarchical  clustering is
expected to occur from bottom up. 
The fact that  low-mass dark haloes are found  statistically to be the
first to 
assemble, has been confirmed both in $N$-body simulations or through the
EPS formalism \citep[e.g.][]{1993MNRAS.262..627L,1997MNRAS.292..835R}.

Their growth is quantified by the mass of the main progenitor, or main
sub-halo that will later merge into it.  
Their epoch of formation can then be defined as the time at which the
mass of the main 
progenitor is half the present mass of the halo.  
However, it is possible to find  the opposite trend, when the number of
merged   haloes  more   massive   than  a   fixed  $M_\text{min}$   is
considered. 
More massive haloes have indeed already assembled most of
their mass in terms of substructures more massive than $M_\text{min}$,
while  less massive  structures are  less advanced  at the  same epoch
\citep{1991MNRAS.248..332B,2006MNRAS.372..933N}. 
This trend can  be called downsizing, which is  similarly observed for
the dark matter, when a minimum mass $M_\text{min}$ is introduced.  

\smallskip

Interestingly, and  for other reasons, the downsizing  observed in the
baryonic  galaxies   is  also  explained   when  a  halo-mass  floor
$M_\text{min}$,  below  which  accretion  is quenched,  is  introduced
\citep{2009MNRAS.397..299M, 2010ApJ...718.1001B}.  
The  cold gas accretion  would then  be limited  to a  particular mass
interval, the  maximum mass  being reached when  the infalling  gas is
sufficiently shock-heated.  
Star formation, linked essentially to the  cold
gas  available for  accretion, then  occurs only  in this  narrow mass
interval.  
At early times, the more  massive haloes reach this floor earlier, and
the mass interval is then crossed more rapidly.  
Star formation in  massive galaxies occurs earlier and  over a shorter
period of time, as the abundance of their elements indicates.
The value of $M_\text{min}$ required to account for observations is of
the order of $10^{11}$~\Msun\ \citep{2010ApJ...718.1001B}.  
Gas exhaustion can then account  for any decline in the star formation
activity \citep[e.g.][]{2007ApJ...660L..47N}. 
However, no  physical process can  explain the existence of  this mass
floor. 
Another  possible explanation could be that  stars form rapidly
in low-mass  haloes early in the  universe, and that most  of these small
galaxies then merge to form  more massive ones, which are observed to
be passively evolving on the red sequence today.  
In this  paper, we wish to  test this possible  scenario, by analysing
$N$-body hydrodynamical multi-zoom simulations.

\smallskip

The  numerical techniques and  the simulation  used are  described in
\S~\ref{simu}.  The derivations of  the bound  structure for  both dark
matter and baryons are presented in \S~\ref{tree}.  
In \S~\ref{resu}, we show physical results of our simulation. 
\S~\ref{acc} describes the results in terms of the fraction of mass
accreted  by galaxies from  the smooth  component versus  mergers, the
influence of the environment being detailed.  
Our considerations of downsizing are presented in \S~\ref{sec:downsiz}.
The discussion in \S~\ref{discuss}  compares the various scenarios for
explaining the downsizing process, and our conclusions are drawn in
\S~\ref{conclu}.

\section{Simulations}

\subsection{Techniques} 

We   use     a   multi-zoom   simulation   based   on  a   TreeSPH
code~\citep{1989ApJS...70..419H}    that     is     described     in
\citet{2002A&A...388..826S}.  
Simulating  galaxy formation  in a  cosmological framework  involves a
wide dynamical range,   to simulate a  large enough simulation
box and to reach a high resolution. 
We    use    here    the    multi-zoom    technique    described    in
\citetalias{2005A&A...441...55S}.  
The  cosmological parameters used  in this  simulation are  taken from
WMAP~3 results
$(\Omega_\text{b},\Omega_\text{m},\Omega_\Lambda,h,\sigma_8,n)        =
(0.042,0.24,0.76,0.73,0.75,0.95)$. 
We  start from an  initial low-resolution  simulation, referred  to as
the ``level 0 zoom'', consisting of $128^3/2$ DM particles and $128^3/2$ gas 
particles, which has mass resolutions of $m_{0,\text{DM}}\simeq 7.2\times
10^{10}$~\Msun{}  and  $m_{0,\text{b}}\simeq 1.54\times  10^{10}$~\Msun{}
and a force resolution $\eps =  50$~kpc, in a cubic box of length $L_0
= 137$~Mpc (comoving).  
We
resimulate a sub-region of the original volume at a higher resolution.  
Tidal  fields and  the inflow  of particles  into and  the  outflow of
particles  away from  the region  of  interest are  recorded at  every
timestep, and the resimulation is run with these boundary conditions. 
At  each level,  we  zoom  in by  a  factor of  two,  improving the  mass
resolution by  a factor  of eight, i.e.  a low resolution  particle at
level $N-1$ zoom 
becomes eight high resolution particles at level $N$.
This  technique  enables any  shape  of  zoom  region, and  we  chose
spherical regions. 
We used three levels of zoom, and ended with a spherical region of radius
$R_3 =  8.56$~Mpc, a mass  resolution of $m_{3,\text{b}} =  3.01 \times
10^7  $~\Msun{}, $m_{3,\text{DM}}  = 1.42\times  10^8 $~\Msun{},  and a
force resolution $\eps = 6.25$~kpc (comoving). 

\smallskip

\begin{figure}[t]
  \begin{center}
    \includegraphics[width=8cm]{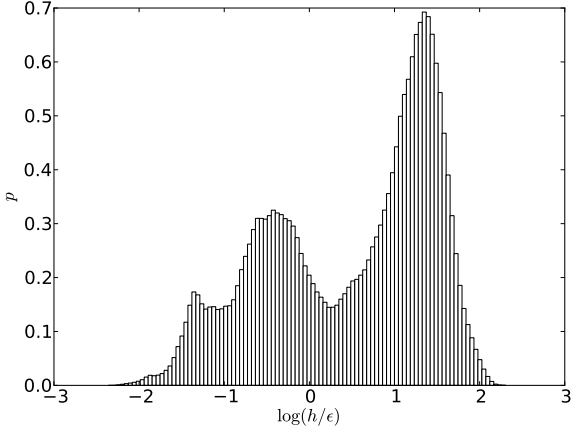}
    \caption{  Distribution  of  the  SPH smoothing  length  $h$  with
      respect to the gravitational  softening parameter $\eps$ at $t =
      9$~Gyr in the level 3 zoom. 
      \label{fig:hist_h}
    }
  \end{center}
\end{figure} 
 
These $128^3/2$ particles of  mass $m_{0\text{DM}}$ and $m_{0\text b}$
at the level~0 give the same mass resolution as a simulation of $128^3\times 
8^3/2 = 1024^3/2$ particles  of mass $m_{3\text{DM}} \simeq 1.42\times
10^{8}$~\Msun{} and $m_{3\text b} 
\simeq 3.01\times 10^{7}$~\Msun{}, but focused on a smaller volume of
radius 8.56~Mpc at level 3. 
This technique enable us to simulate galaxies with a fairly high resolution
starting with a reasonably sized cosmological box, at a smaller CPU cost. 

\smallskip

One of the  main characteristics of this technique  is that, except at
the  zeroth level of  zoom, the  number of  particles does  not remain
constant during the whole simulation.  
At higher levels of zoom, a  particle at timestep $i$ inside the level
$N$ box, but  outside the level $N+1$ box, can  indeed enter the level
$N+1$ box  at timestep $i+1$  and be split into  eight high-resolution
particles.  
Particles with  unrecorded history  the enter the  box, and  a special
care must be taken when establishing particle identities.
Because of  this increasing  number of particles,  the third  level of
simulations could not be run further than $t=9.1$~Gyr, or $z =
0.46$, but lower levels of zoom reached $z= 0$.

\smallskip

At the third level of zoom, we end up with 90 snapshots,
  sampled  every  100~Myr  from  $t=0.2$~Gyr  to  $t=9.1$~Gyr,  which
  enables  us to build  the merger  tree of  structures, while  at the
  three  lower levels  of  zoom  we have  70  snapshots, sampled  every
  200~Myr from $t=0.2$~Gyr to $t = 14$~Gyr.
The   latter   can   be   used   for   a   resolution   study   (see
section~\ref{sect:resol}). 
The   properties  of   each   level   of  zoom   are   summed  up   in
table~\ref{tab:sim_zoom}.

\subsection{Physical recipes and initial conditions} 
\label{sec:ic}

While collisionless  particles, namely stars and  dark matter, undergo
only gravitational  forces and  are treated by  a tree  algorithm, gas
dynamics is treated by smooth particle hydrodynamics (SPH). 
Additional recipes  are needed to  mimic subgrid physics such  as star
formation and feedback. 
Our physical treatment is described in \citet{2002A&A...388..826S}, 
{but the present paper we only use the SPH phase and not the
  cold and clumpy gas that was described by sticky particles.  } 
The SPH  gas is treated with the  same equation of state  and the same
viscosity prescription.  
The range of temperatures is $800-2\times 10^6$~K.
Any dependance of cooling on metallicity is ignored, and only a primordial
metallicity ($10^{-3}$~Z$_\odot$) is considered.  
A unique timestep per zoom level is adopted, which is respectively 20,
10, 5, and 2.5~Myr for levels 0 to 3. 

The softening  length for  gravity is respectively  50, 25,  12.5, and
6.25~kpc (comoving), and we checked that the SPH smoothing length is 
not far shorter than the softening length, as shown
on the histogram of Fig.~\ref{fig:hist_h}  at $t = 9$~Gyr in the level
3 zoom.

\begin{figure}[t!]
  \begin{center}
    \includegraphics[width=8cm]{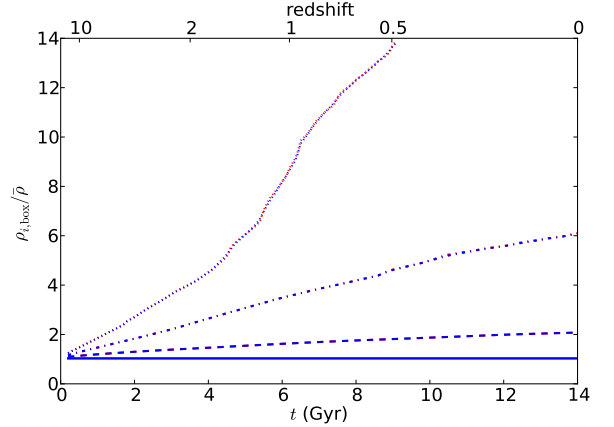}
    \caption{ Evolution of  the mean comoving density of  the box with
      respect to the cosmic density
      $\rho_{i,\text{box}}(t)/(\bar\rho_i)$   for  the   four  zoom
      levels. Level 0 is shown as a solid line, level 1 a dashed line,
      level 2 dash-dotted and level 3 dotted.  
      The box in the last snapshot a level 3 has a density of 14 times
      the cosmic density for $i\in$ (DM, baryons). 
      \label{fig:rhom}
    }
  \end{center}
\end{figure} 

The  radiative cooling  term $\Lambda$  is taken  from  the normalised
tables of \citet{1993ApJS...88..253S} modelling atomic 
absorption-line cooling from $10^4$~K to $10^{8.5}$~K.
The  background ultraviolet  (UV)  radiation field  is  modelled by  a
constant uniform heating $\Gamma_\text{UV} = 10^{-24}$ erg s$^{-1}$ term.

\begin{figure*}[ht!]
  \begin{center}
    \includegraphics[width=17cm]{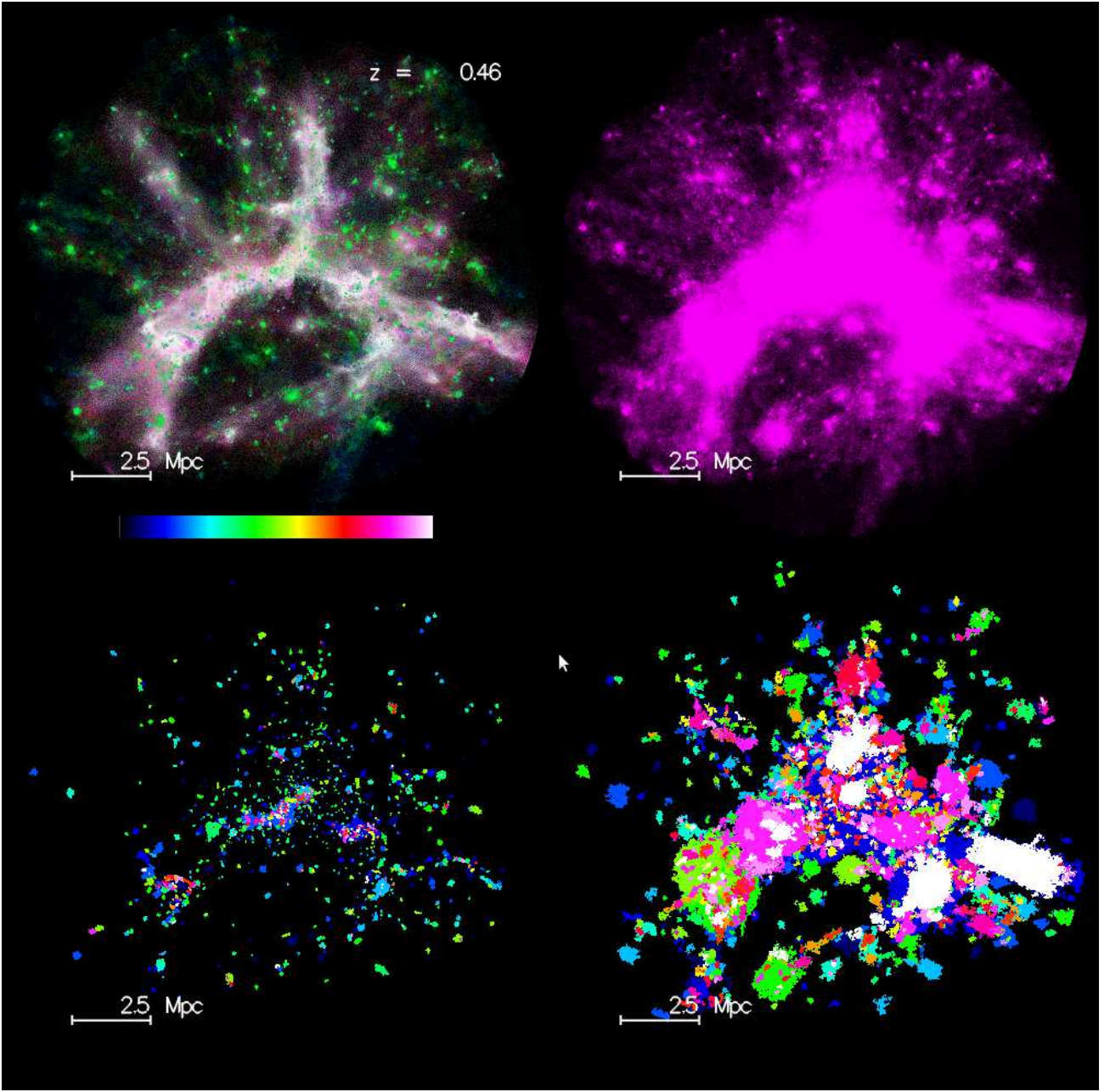}
    \caption{View of the third zoom level of the simulation.
      \emph{Top}: Gas  (colour-coded by temperature,  logarithmic scale
      from 800 to 
      $1.4\times 10^6$~K), and DM (right). \emph{Bottom:}
      Structures and  substructures detected by  AdaptaHOP. \emph{Left}:
      baryonic  galaxies  and satellites,  \emph{right}:  DM haloes  and
      subhaloes. 
      Haloes and galaxies are represented  in dark and bright blue and
      green; subhaloes and satellites in yellow, orange, magenta, red,
      and white. 
      The white bar indicates the comoving length-scale.
      \label{fig:simu}
    }
  \end{center}
\end{figure*}

Star formation is modelled by a  Schmidt law with a star formation rate
of 
\begin{equation}
  \label{eq:sf}
  \deriv {\rho_*} t = C \rho_\text{gas}^n,
\end{equation}
with $n = 1$. 
It  is applied to  gas particles  with densities  higher than  a density
threshold of 
\begin{equation}
  \label{eq:thr}
  \rho_\text{min} =3 \times 10^{-2}\,\text{at cm}^{-3}.
\end{equation}
Gas particles form stars, and have a fraction of stars within them. 
When this fraction reaches a given threshold (set to 20\%), we
search  among their neighbours  to determine  whether there  is enough
material to form 
a  full  star particle,  \emph{i.e.}   whether  the  sum of  the  star
fractions among the neighbouring gas particles is greater than 1.  
If this  is the case, the particle  is turned into a  star particle, and
the fraction of stars in the neighbouring particles becomes gas again.
In this way,  the stellar fraction within a gas  particle remains low,
which prevents stars from following the gas dynamics.
Kinetic feedback from supernovae (SNe) is also included. 
Stars more massive  than 8~\Msun{} are assumed to  die as SNe,
releasing an energy of $10^{48}$~erg~\Msun$^{-1}$. 
The released  energy is distributed  assuming a Salpeter  initial mass
function, and an 
efficiency parameter of 6\%. 
Particles  within the  smoothing  length of  the  former gas  particle
receive a velocity kick in the radial direction.

We note that in this study, no AGN feedback is considered, which
  leads to an overcooling problem and to high stellar masses in the more
  massive haloes (see \S~\ref{sec:agn}).

\smallskip

{Initial    conditions    were    obtained   using    Grafic
  \citep{2001ApJS..137....1B}  at  the  highest  resolution  ($1024^3$
  particles)  for level  3, and  were then  undersampled to  build the
  initial conditions of low-resolution levels. } 

We ran the level 0 simulation and ran a FOF-like algorithm to detect
the  haloes to be resimulated. 
We chose the most massive  halo of about $10^{15}\Msun$ at $z=0$, thus
all this work is done in a rich environment. 
Figure~\ref{fig:rhom}  shows  the   evolution  of  the  mean  density,
normalised by the cosmic density, for each zoom level.  
The dotted line is level 0,  the dash-dotted level 1, the dashed level
2, and the solid level 3.

\section{Building merger trees} 
\label{tree}

\begin{figure}[t!]
  \begin{center}
    \includegraphics[width=8.5cm]{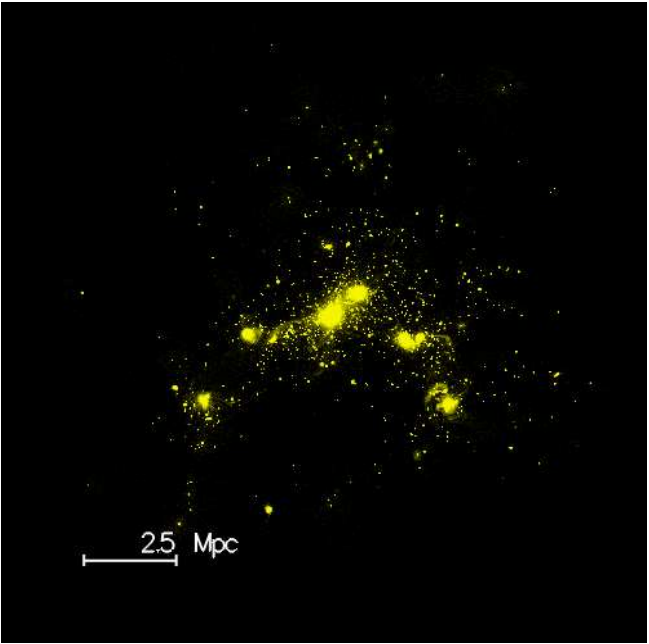}
    \caption{View of the third zoom level of the simulation (continued):
      stars. 
      \label{fig:simu_stars}
    }
  \end{center}
\end{figure}

Deriving merger trees  from $N$-body simulations is not  an easy task,
because there is  still a lot of freedom in  defining halo and subhalo
masses, and identifying both progenitors and sons in merger trees.  
There are  two essential  steps building  merger trees:  detecting the
structures, and linking them from one timestep to the other.
A   ``bad''  structure  detection   {results  into}   a  bad
definition  of the  detected structures,  and makes  it  impossible to
extract results. 

\smallskip

We  used here  the  AdaptaHOP algorithm,  and  we followed  the rules  of
\citetalias{2009A&A...506..647T}  defining  the  haloes and  subhaloes
hierarchy as well as the merger history.

\smallskip

In addition, since we aim to study galaxies, we wish to detect
separately the  baryonic components,  namely the central  galaxies and
their satellites. 

\subsection{Structure detection}

Structure  finders  are  widely  used in  computational  cosmology  to
analyse simulations. 
Structure  finders   of  the  first   generation,  such  as   the  FOF
algorithm~\citep{1985ApJ...292..371D}, were  able to detect virialised
DM haloes in the simulations. 

\smallskip

{The  FOF  algorithm links  together  particles closer  than
  $b_\text{link}$ times the mean interparticle distance. 
It  is efficient  in finding  haloes, but  tends  to  link together
separate objects when they are too close to each other, especially
during mergers.}

\smallskip

Spherical  overdensities~\citep[SO,  ][]{1994MNRAS.271..676L}  follows
another  approach: it  detects  density maxima,  and  grows from  each
maximum a sphere such that 
the mean  density within this  sphere is equal  to a given  value, e.g.
$\Delta_c  \times\rho_c$,   where  $\Delta_c  \simeq   200$  (slightly
depending on the cosmology and the redshift) is the virial
overdensity, and $\rho_c$ is the critical density of the universe.

\smallskip

\textsc{Denmax}~\citep{1994ApJ...436..467G},     Bound    Density
Maximum~\citep[BDM,][]{1999ApJ...516..530K}                         and
SKID~\citep{2001PhDT........21S} compute the density field, and move 
particles along the density  gradients to find local maxima. 
\textsc{Denmax}  computes the  density on  a grid,  while in  SKID the
density field and its gradient are computed in a SPH way.
Unbound particles are then iteratively removed via an unbinding step.
HOP~\citep{1998ApJ...498..137E} has  a similar spirit,  but instead of
computing density gradients, it  jumps {from one particle to
  its denstest  neighbour, thus efficiently  indentifies local density
  maxima}.  
SUBFIND \citep{2001MNRAS.328..726S} finds  subhaloes within FOF haloes
by  identifying saddle  points.   The  density is  computed  in a  SPH
fashion. 
Each particle has a list of its two densest neighbours.  
Particles   without  denser  neighbours   are  density   maxima,  thus
correspond to a new substructure. 
Substructures then grow towards lower density particles. 
Saddle points  are defined as  particles that have their  two densest
neighbours belonging to two different substructures. 
AdaptaHOP \citep{2004MNRAS.352..376A} is close to SUBFIND, except that
the construction of the substructures occurs in a bottom-up manner, as
we will later describe.  
\textsc{Amiga}'s  Halo  finder \citep[AHF,][]{2009ApJS..182..608K}  is
quite similar, but computes the  density on an adaptive mesh refinement
(AMR) grid, taking advantage of the adaptive nature of the AMR.
VOBOZ~\citep{2005MNRAS.356.1222N} is again  similar, but uses a Voronoi
tessellation to estimate the density.
 PBS \citep{2006ApJ...639..600K} uses a total boundness criterion
combined  with  a tidal  radius  to  define  substructures within  FOF
haloes.

\begin{figure*}[ht!]
  \begin{center}
    \includegraphics[width=17cm]{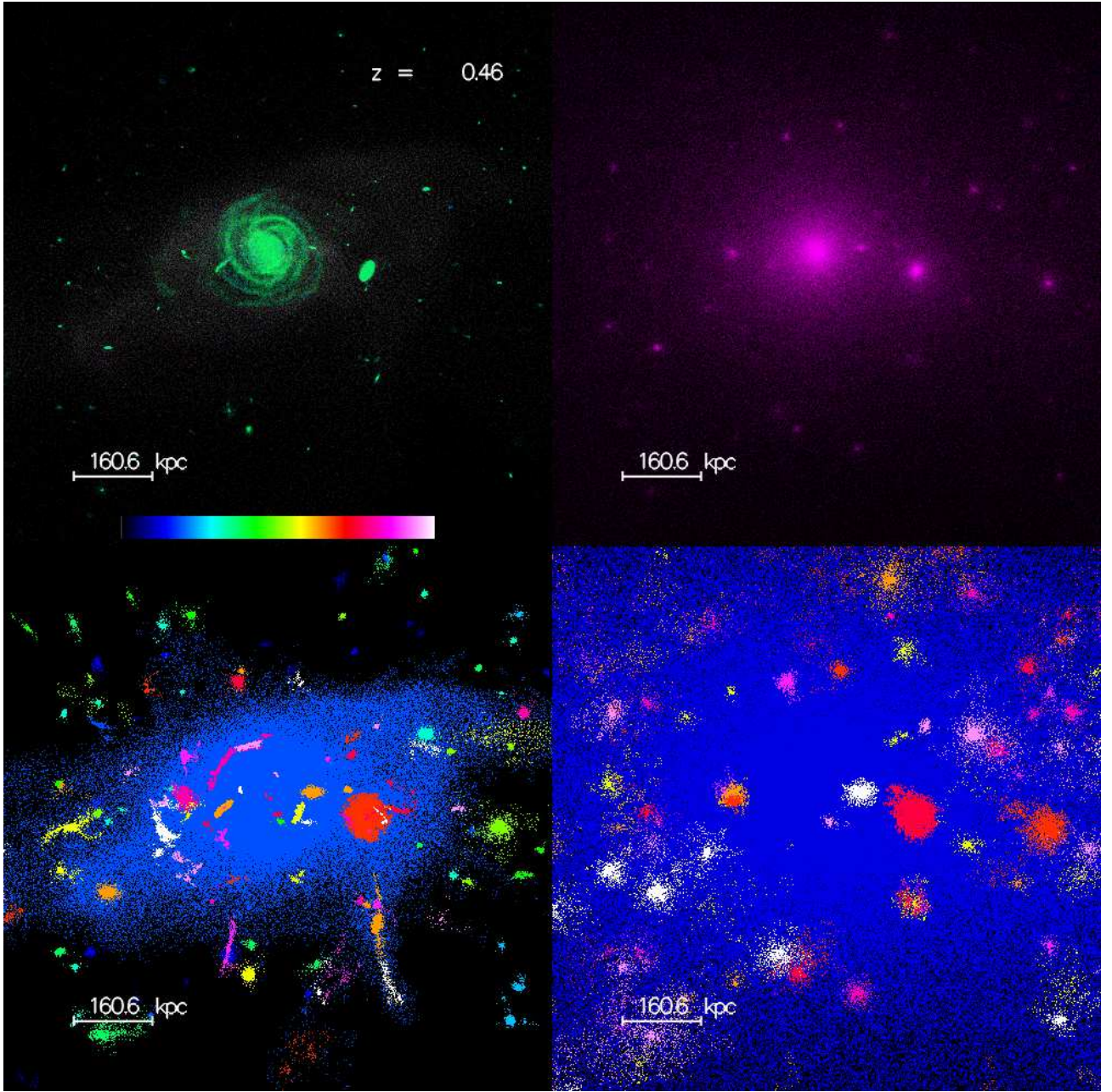}
    \caption{Zoom  on the  most massive  halo  of the  third level  of
      zoom. 
      Same legend as figure \ref{fig:simu}.
      \label{fig:clus}
    }
  \end{center}
\end{figure*}

\smallskip

All those algorithms are however only 3D, and based on the position of
the particles.  
Several algorithms  take advantage of the availability  of the velocit
data in addition to the positions. 
Six-dimensional (6D) FOF~\citep{2006ApJ...649....1D} is a FOF algorthm
with a 6D metric.  
Hierarchical  structure finder~\citep{2009MNRAS.396.1329M} works  in a
similar way to SUBFIND in 6D, using a 6D density estimator.  
\texttt{Rockstar} \citep{2011arXiv1110.4372B} uses a 6D metric with
the addition of time information

\smallskip

For a recent and more detailed comparative study of halo finders, we
refer to \citet{2011MNRAS.415.2293K}. 
The structure finder must be applied to every snapshot of the simulation
in order to build a merger tree. 

We  used AdaptaHOP to  detect both  DM and  baryonic structures  in the
simulation. 
AdaptaHOP proceeds in four steps:
\begin{itemize}
\item First,the SPH density of particles is computed over $N_\text{ngb}$ 
neighbours.  We chose $N_\text{ngb} = 32$. 
\smallskip

\item Then, all particles whith density higher than the
  user-defined threshold density \rt{} are selected, and the algorithm
  then jumps (``hop'') to their 
  densest neighbour, thus finds the local maximum they belong to.  
  A  tree  of  structures  is  then  built, the  leaves  of  the  tree
  corresponding to particles belonging to the same local maximum.

\item Finally, the saddle points that link together two structures are
  identified.

\item The algorithm then reiterates  within each leaf of the structure
  tree to detect substructures. 
\end{itemize}

The result is a tree of structures and substructures, where the leaves
correspond  to physical  structures.
The main parameters of AdaptaHOP are: 
\begin{itemize}
\item   $N_\text{members}$:  minimal   number  of   particles   for  a
  (sub)structure to be considered. 
  It gives the minimal mass for a (sub)structure.
\item  \rt:density threshold  of the  first  level, which  is used  to
  detect main haloes.  
  Particles with a SPH density  below \rt are part of the
  background. 
\item {$\alpha$: peak of the substructure. Only substructures with a density maximum $\rho_\text{max} > \alpha
    \bar\rho_\text{sub}$ are considered significant. }
\item $f_\text{p}$: Poisson noise parameter.  
  The existence of  a substructure is tested by  comparing its density
  wih the Poisson noise: 
  a substructure with $\bar\rho_\text{sub} > \rho_\text{s}\left(1+\frac
    {f_\text{p}}{\sqrt N}\right)$ is statistically significant, where
  $\rho_\text{s}$ is  the density of the saddle  point that separates
  the substructure from others, $\bar\rho_\text{sub}$ the mean density
  of the  substructure, and $N$  the number of particles  belonging to
  the substructure.  
\item $f_\eps$: controls the size of the structure.  
  Every structure must have a radius larger than $f_\eps$ times the mean
  interparticle distance.
\end{itemize}

Following  \citetalias{2005MNRAS.363....2K}, we  set  $N_\text{members} =  64$,
which yields a minimum mass of 
$ \simeq 8.96\times 10^9$~\Msun{} for theDM haloes, and $1.92 \times 10^9
$~\Msun{} 
for baryonic structures.

\smallskip

We note that we had to slightly modify the algorithm to make it
compatible with the multi-zoom technique.
Since particles enter the box between successive
timesteps, $\bar\rho_\text{box}$ is indeed not constant, as opposed to
regular simulations, and the evolution of $\bar\rho_\text{box}(t)$ for
the four levels of zoom
is shown in figure~\ref{fig:rhom}.  
We thus compared the density to \rt times the mean density in
the level 0 zoom, since the mean density in higher levels is higher
than the mean density of the universe.
At the last output, the mean density in the third level of zoom that we
analyse is about 14 times the cosmic density.

\smallskip

\subsubsection{DM halo detection}

\begin{figure}[t]
  \begin{center}
    \includegraphics[width=8.5cm]{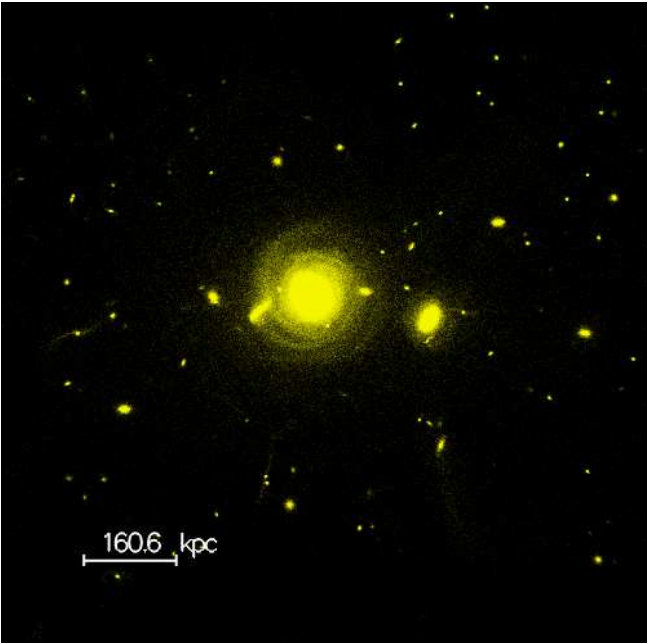}
    \caption{Zoom on the most massive  halo of the third level of zoom
      (continued): stars.
      \label{fig:clus_stars}
    }
  \end{center}
\end{figure}

Following~\citet{2004MNRAS.352..376A}
and~\citetalias{2009A&A...506..647T}, we defined haloes as groups of 
particles  with densities  higher than  a given  threshold  density, and
subhaloes as locally  overdense groups of particles within  a host (or
main) halo, which separed by density saddle points.  
The centres  of the haloes were  defined as the positions  of the densest
particles in the halo rather than the centres of mass. 
The reason for  this choice is that for a major  merger, the centre of
mass can be halfway between the two merging objects, and we preferred 
to define the ``real centre'' as the centre of the main halo. 

{In the following, the mass of a halo is defined as the mass
  of the main halo plus the  mass of the subhaloes, therefore the mass
  of a subhalo can be counted several times.}

As   advocated   in~\citet{1998ApJ...498..137E},   $\rt   =
80\times  \bar\rho$,  where $\bar\rho$  is  the  mean  density of  the
universe  (see previous  section),  was  set for  DM  haloes, which  is
roughly equivalent to a linking length $b_\text{link}=0.2$ in FOF.  
We  kept default  values for  the other  parameters: $f_\text{P}  = 3,
\alpha = 1, f_\eps = 0.05$.

\subsubsection{Galaxy detection}

Several attempts have been made to build baryonic merger trees.
\citet{2007MNRAS.377....2M}  used  SKID~\citep{2001PhDT........21S}  to
detect baryonic  structures, and they referred to  as ``family trees''
the merger trees of baryonic galaxies, in order to avoid any confusion
with subhaloes merger trees.
However, they  have only considerred  star particles, whereas  we also
wish to take into account gas.
Other authors have used SKID to detect both stars and cold gas, with
$\rho/\bar{\rho}    >    1000$   and    $T    <   3\times    10^{4}$~K
\citep[e.g.][]{2009MNRAS.399..650S}.

\smallskip

\citet{2006ApJ...647..763M}  also  used  SKID and  introduced  ``virtual
galaxies''  in order  to  account for  the  fragmentation of  baryonic
structure between consecutive outputs.
On  the  other  hand,  \citet{2009MNRAS.399..497D,2010MNRAS.405.1544D}
used a modified  version of SUBFIND  to detect simultaneously  dark matter
and baryons, ending with a galaxy composed of DM, stars, and gas. 
They used dynamical criteria  to distinguish between the central galaxy
and the diffuse stellar component. 

\smallskip

Our approach here was slightly  different: we detected on the one hand
the hierarchy  of DM haloes and  subhaloes, and on the  other hand the
baryonic component consisting of central galaxies and satellites.

We therefore also used AdaptaHOP to detect baryonic structures.
While input  parameters are known for  DM detection, we had  to find a
more well-suited set of parameters in order to detect the baryonic
structures. 
 We set the density threshold \rt{} above which structures are 
considered to 1000 times the mean (baryonic) density.
We took  $f_p = 4$, $f_\eps  = 5\times 10^{-4}$
in  order to  allow to  detect structures  with sizes  of the  order of
1~kpc, and kept $\alpha$ = 1.
Higher  values of  $f_p$ tend  to remove  the  smallest structures,
while lower values add unphysical substructures.

\subsubsection{Matching dark and baryonic structures}
\label{sec:matching}

An interesting question that we  addressed is how much baryonic matter
there is in DM haloes.  
To  answer this  question, we  studied the  link between  galaxies and
haloes, taking advantage of their independent detections.  
We then  set several rules to decide  whether a galaxy and  a halo are
linked together.  

\smallskip

The first rule is that a galaxy should belong to at most one (sub)halo
(and of course its host halo hierarchy if it is a subhalo).  
The second  rule is that the  hierarchy of galaxies  and satellites on
the  one  hand  and haloes  and  subhaloes  on  the  other has  to  be
respected, so  that we avoid the  case where a satellite  is linked to
the host halo while the galaxy is linked to the subhalo. 
With these rules, a halo $h$ can host several galaxies,
but at most one main galaxy $g$, which is the most massive
galaxy of $h$ and must have $h$ as its halo. 
Bearing these rules in mind, we were able to match DM haloes to galaxies.

\subsection{Merger tree building}

In  the context  of  structure  finding, one  of  the most  persistent
problems is the so-called flyby issue.  
This phenomenon can  occur when two haloes cross  each other, and when
their respective centres are too close to each other.
They are then detected as only  one halo -- even if they can sometimes
still be
distinguished by eye -- and thus considered as a merger, but
detected again as two separated haloes at a later timestep.  
\smallskip

This problem can be partially  resolved by using a subhalo finder such
as AdaptaHOP instead of a halo finder such as FOF, since the second
halo can  still be tracked as  a subhalo, thus  conserve its identity;
however the problem remains when the  subhalo is too close to the host
halo centre.  
An improvement would be to use a phase-space halo finder, such as
HSF \citep[cf][]{2009MNRAS.393..703M,2009MNRAS.396.1329M}. 
\citet{2012ApJ...751...17S}  introduced an  interesting  method called
``halo interaction network'', which is a more complex 
merger tree that takes into account flybies. 
However,  for  this work  we  were  only  interested in  discriminating
between particles entering smoothly from the background and particles
belonging to another  structure and entering by means  of mergers, hence
we did not need such a refinement.

\citet{2009A&A...506..647T} give different sets of rules building 
DM merger trees that include subhaloes, where a (sub)halo at output
$t_{n+1}$ is the son of its progenitor at output $t_n$. 
We refer to a structure as either a subhalo or a halo.
These rules are:

\begin{itemize}
\item A structure can have at most one son.
\item The  son of structure $i$ at  output $t_n$  is the structure  $j$ at
  output $t_{n+1}$, which inherits most of the mass of structure $i$.
\item Structure $i$ at  output $t_n$  is a progenitor of structure $j$  at
  output $t_{n+1}$, if $j$ is the son of $i$. 
\end{itemize}

\begin{figure*}[t]
  \begin{center}
    \subfigure[face on]{
      \includegraphics[width=17cm]{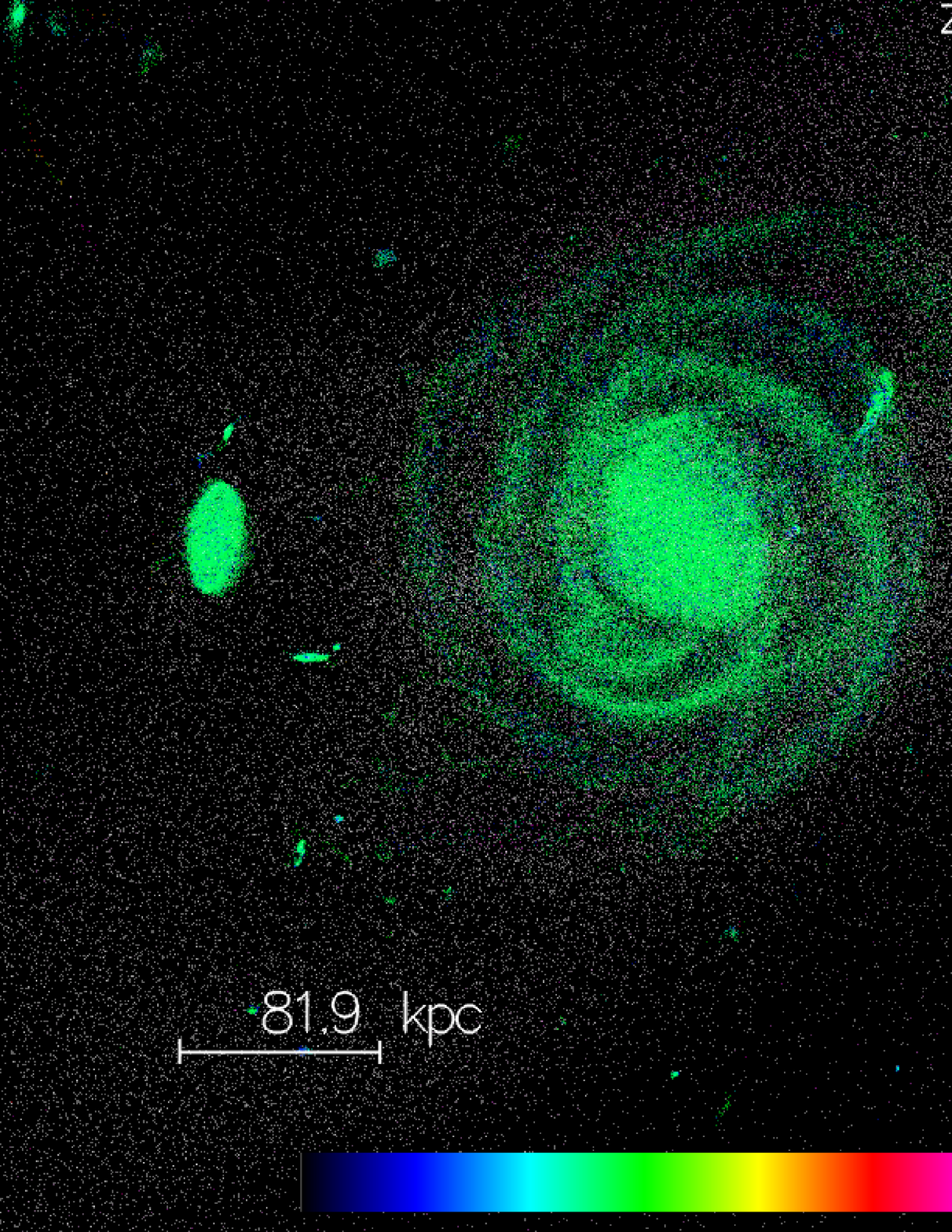}
      \label{fig:faceon}
    }
    \subfigure[edge on]{
      \includegraphics[width=17cm]{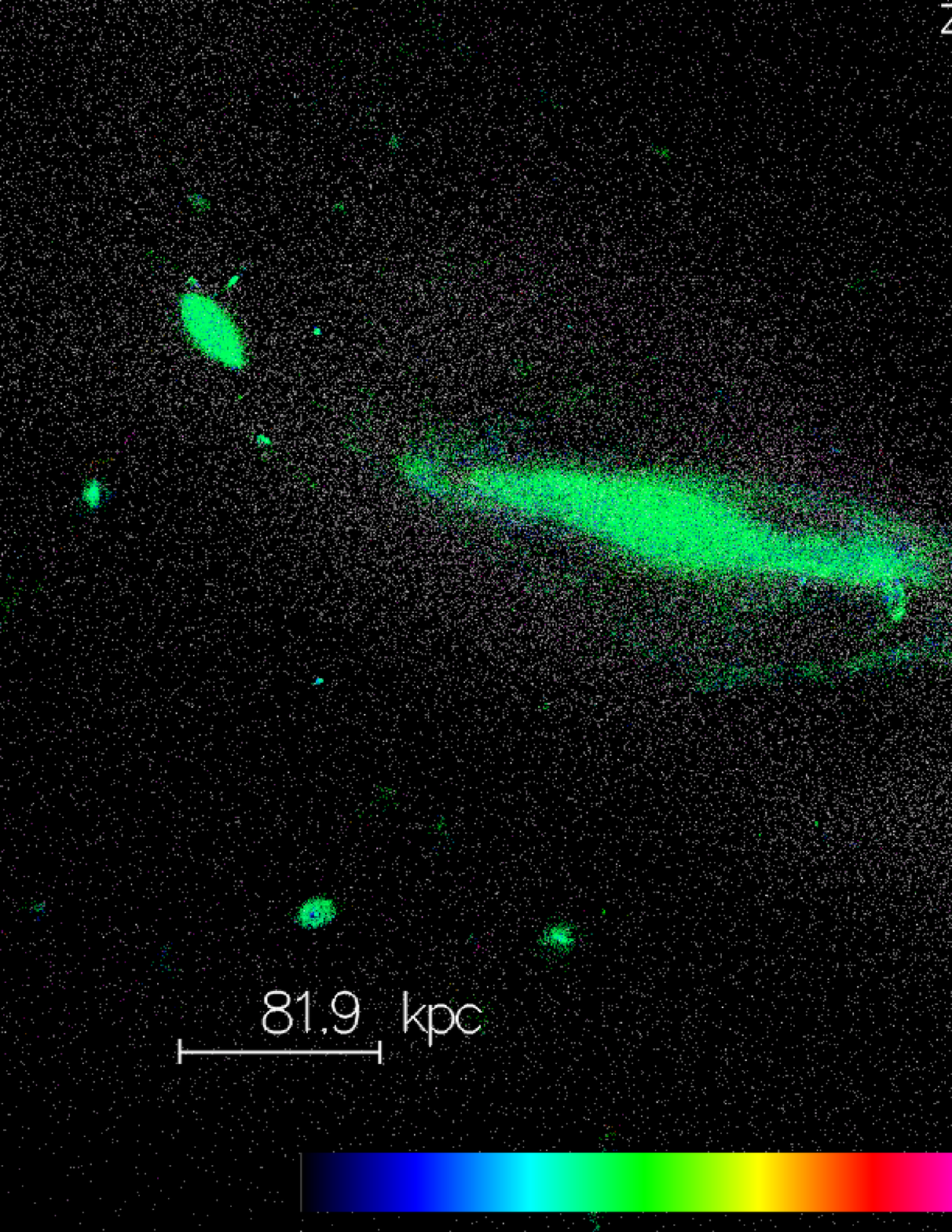}
      \label{fig:edgeeon}
    }
    \caption{
      View of the most massive  central galaxy, at $t=9.1$~Gyr, in the
      third level of zoom. Left: gas colour-coded
      by temperature from 800 to $1.4 \times 10^{6}\,\text{K}$, right:
      stars.
      \label{fig:central}
    }
  \end{center}
\end{figure*}

They introduced a two-step method called the branch history method (BHM)
to determine which of two local  maxima should be the subhalo
and which the main halo, according to the results of the previous step. 

This method tends to avoid  identity switches between the main halo
and the satellite. 
The basic idea is to take advantage of the previous snapshot to decide
which node should be the subhalo and which the halo.

\smallskip

Once again,  we had to modify  the algorithm to take  into account the
number of particles, which is not  constant with time owing to our use
of the multi-zoom method.

\subsection{Accretion history}

After we had  built the full merger tree of each  galaxy, we were able
to compute the mass history.  
We traced back the main progenitor from the last snapshot to the first,
and  tagged  each  particle  entering  the  main  structure  at  each
snapshot. 
Particles coming from either a satellite of the considered galaxy or from
another galaxy were
tagged as  \emph{merger}, while particles coming  from the background
were defined as \emph{smooth accretion}.

\smallskip

Particles can also  leave the main galaxy, either  for the background,
which we refer to \emph{evaporation}, of for another substructure, which
we dub \emph{fragmentation}. 
The latter  happens mainly during mergers events:  particles from a
satellite are detected as part of the main galaxy at a given snapshot,
but may have left before the following. 
The former can happen at almost every snapshot: for particles at the
border of  the structure, density  can fluctuate without  moving, thus
be on either one side of the saddle point or the other.  

\smallskip

With these definitions, we were able to compute the baryonic mass assembly:
\emph{merger} $-$ \emph{fragmentation} and
\emph{accretion} $-$ \emph{evaporation}.  
The mass of a structure was counted only once, since a particle entering the
galaxy is counted positively, and negatively when it leaves.
This enables us to overcome the fly-by issue mentioned above, since
a fly-by would be counted first as a merger, then as fragmentation and could
then vanish in the total accretion fraction.
\smallskip

However, there is another difficulty: as a consequence of the 
multi-zoom technique, particles enter a higher zoom-level box at each 
timestep, thus several galaxies enter the box when they have already
formed. 
Accretion fractions are computed between $t_\text{app}$, the time when
the galaxy enters the last zoom level box, and $t_\text{end}$, the end
of the simulation.  
We thus concentrated  on galaxies entering the box  before $t = 7$~Gyr
in order to follow them over a sufficient number of timesteps.

\section{Results} 
\label{resu}

\subsection{Structure detection and merger trees}

\begin{figure*}[ht!]
  \begin{center}
    \includegraphics[width=17cm]{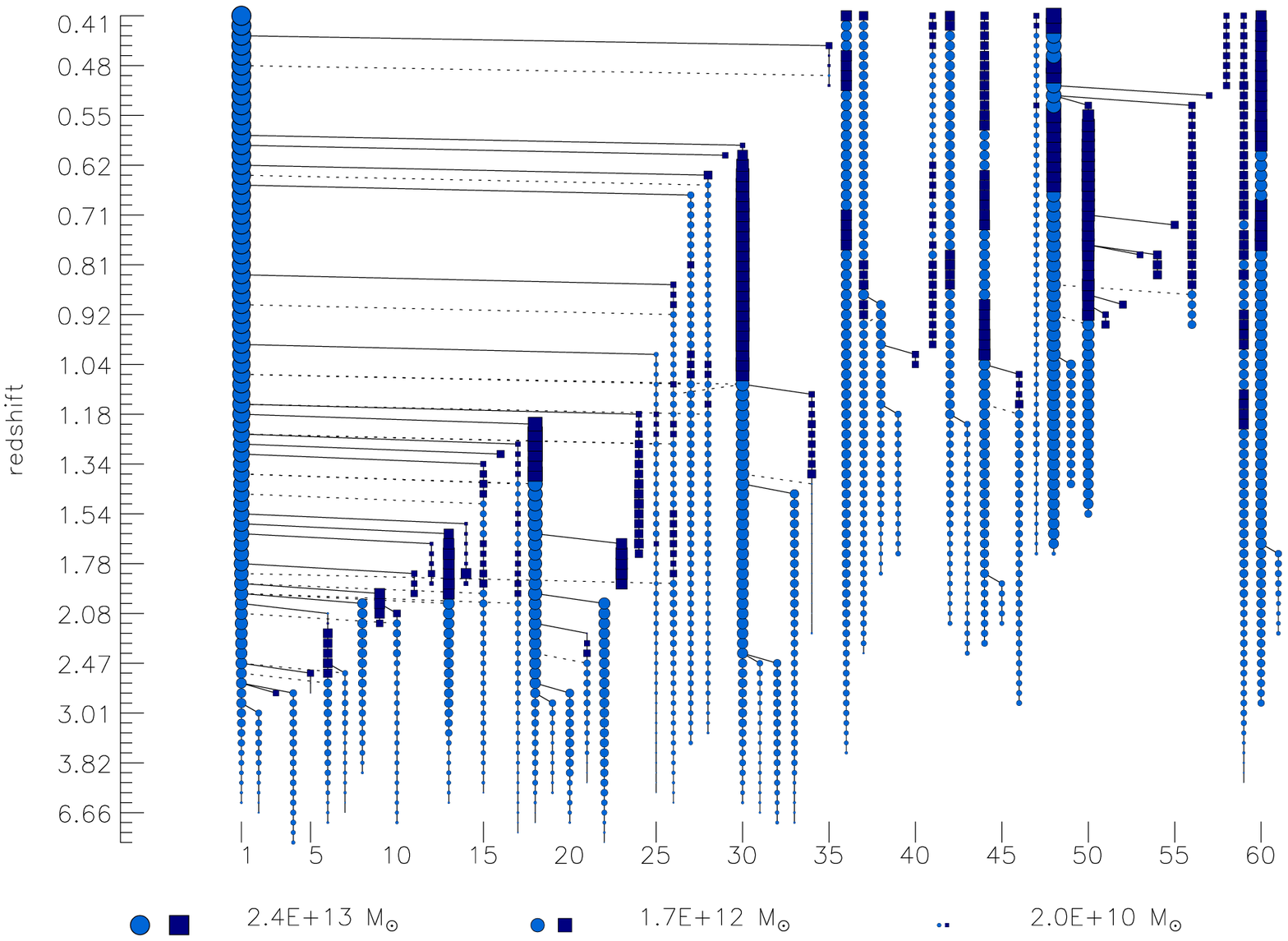}
    \caption{Baryonic merger tree of  the main galaxy. Dark blue circles
      are  galaxies, and bright  blue squares  satellites. The  $x$ axis
      shows the number of the branch, i.e. a galaxy that will eventually
      merge  with the  main galaxy  (branch 1),  or stay  as one  of its
      satellites. 
      \label{fig:tree}
    }
  \end{center}
\end{figure*}

Figures~\ref{fig:simu}   and~\ref{fig:clus}   show,  respectively,   a
large-scale view  of the third level  of zoom of the  simulation and a
zoom on the most massive galaxy of the box.
The upper panel shows   on the left gas
particles  (colour-coded by  temperature, on  a logarithmic  scale from
800~K to $1.4\times 10^6$~K), and on the right, DM
particles.  
Figures~\ref{fig:simu_stars}    and~\ref{fig:clus_stars}    show   the
corresponding star distributions.

\smallskip
The lower  panel shows the structures detected  by AdaptaHOP, baryonic
galaxies and  satellites on the left,  and DM haloes  and subhaloes on
the right.
Haloes and main galaxies appear in  dark and light blue and green, and
subhaloes and  satellites appear in yellow, orange,  red, magenta, and
white. 
At  the centre,  the most  massive  halo (in  blue) can  be seen  with
massive subhaloes  (in magenta  and white): it  is undergoing  a major
merger at this timestep, which explains why these massive substructures
{appear larger than several small and isolated haloes}.

\smallskip

We can see by eye the good agreement between the baryonic and DM
structure detected. However,  since for a given galaxy  the DM halo is
far more  extended than the  baryonic structure, this  is not easy. In
section~\ref{sec:matching}, we  explained how we  matched galaxies and
haloes. 

\smallskip

The zoom in figure~\ref{fig:clus} is instructive. We can still discern
the close correspondence between the dark and baryonic structures, and
most of the 
small structures in the upper panels are indeed detected in the lower
panels. 
However, some  remarkable features  can be found:  in the  bottom left
panel, there is a satellite
(in orange, under  the largest satellite in yellow)  with a tidal tail
(in red), which is detected by  AdaptaHOP as a satellite whose tail is
a ``satellite'' of this satellite.  
This  structure  finder  is  thus  capable  of  detecting  interesting
features.  

Most strikingly, the red arc near the centre of the main galaxy is an
artefact:  it is  not a  satellite, but  rather an  arm of  the spiral
galaxy. 

 \smallskip

In figure~\ref{fig:clus}, we can see the central galaxy of the halo,
which contains a large disc of $\simeq 160$~kpc (comoving) at $z=0.46$. 
Figures~\ref{fig:faceon}  and~\ref{fig:edgeeon} show,  respectively, a
face-on and an edge-on view of the central galaxy.
Looking  at  the  corresponding  galaxy   at  $z=0$  in  our  level  2
simulation,  it appears  that  this  galaxy still  has  a gaseous  and
stellar disc  today, which is quite unexpected  since central galaxies
are supposed to be elliptical.
{
The large  size and  mass ($\simeq 10^{13}\,\Msun$)  of the  galaxy is
probably caused by our not taking into account AGN
feedback in our simulations. 
}

Figure~\ref{fig:tree} shows the merger tree of this galaxy.
Only the 60 most massive branches of the tree are shown here.
The $y$ axis is the redshift,  and each branch is a galaxy that either
completely merges  with the  main galaxy, or  becomes a  satellite of
this galaxy at the last timestep. 
The first  branch on  the left is  the \emph{main  progenitor} branch,
i.e. the ancestors of the main galaxy.  
Branches  2 to  35  are  galaxies (bright  blue  circles) that  became
satellites (dark  blue square) of the  main galaxy, and  merged with it
before the last timestep.  
Branches 36 to 61 are the satellites of the main galaxy at the last
timestep, and their merger trees.

Galaxies that  seem to appear in  the merger tree at  low redshift are
actually entering the level 3 zoom at this time (e.g.  branches
42--51). 
However, satellites
that seem to appear late (e.g. branches 23, 24, 42) correspond to fly-bies:
they have no identifiable progenitor at any previous timestep.

\subsection{Evolution of the mass function}
\label{sec:massfunc}
\begin{figure*}[t!]
  \begin{center}
    \subfigure[Galaxies]{
      \includegraphics[width=8cm]{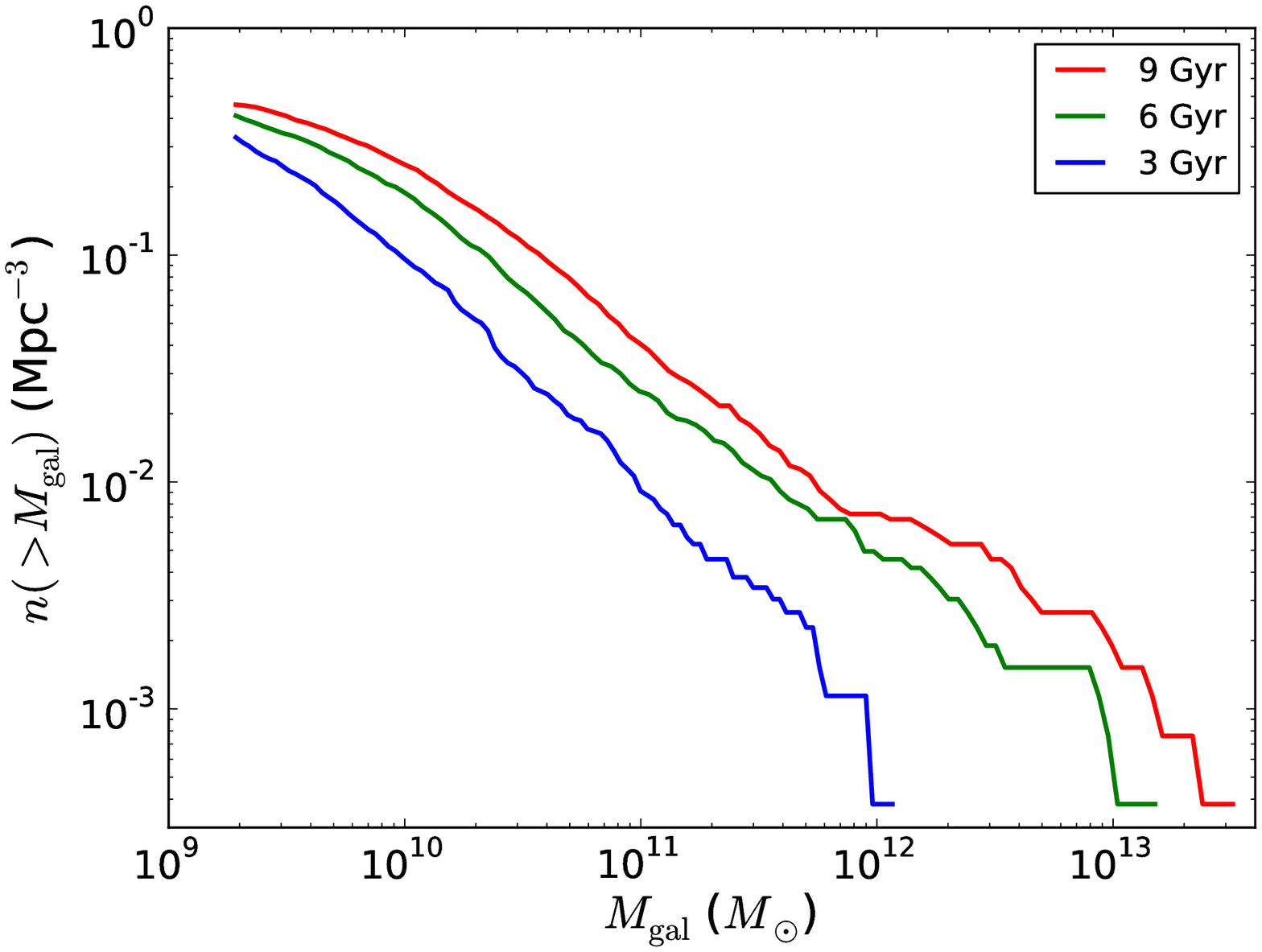}
      \label{fig:mspec_bar} 
    }
    \subfigure[Haloes]{
      \includegraphics[width=8cm]{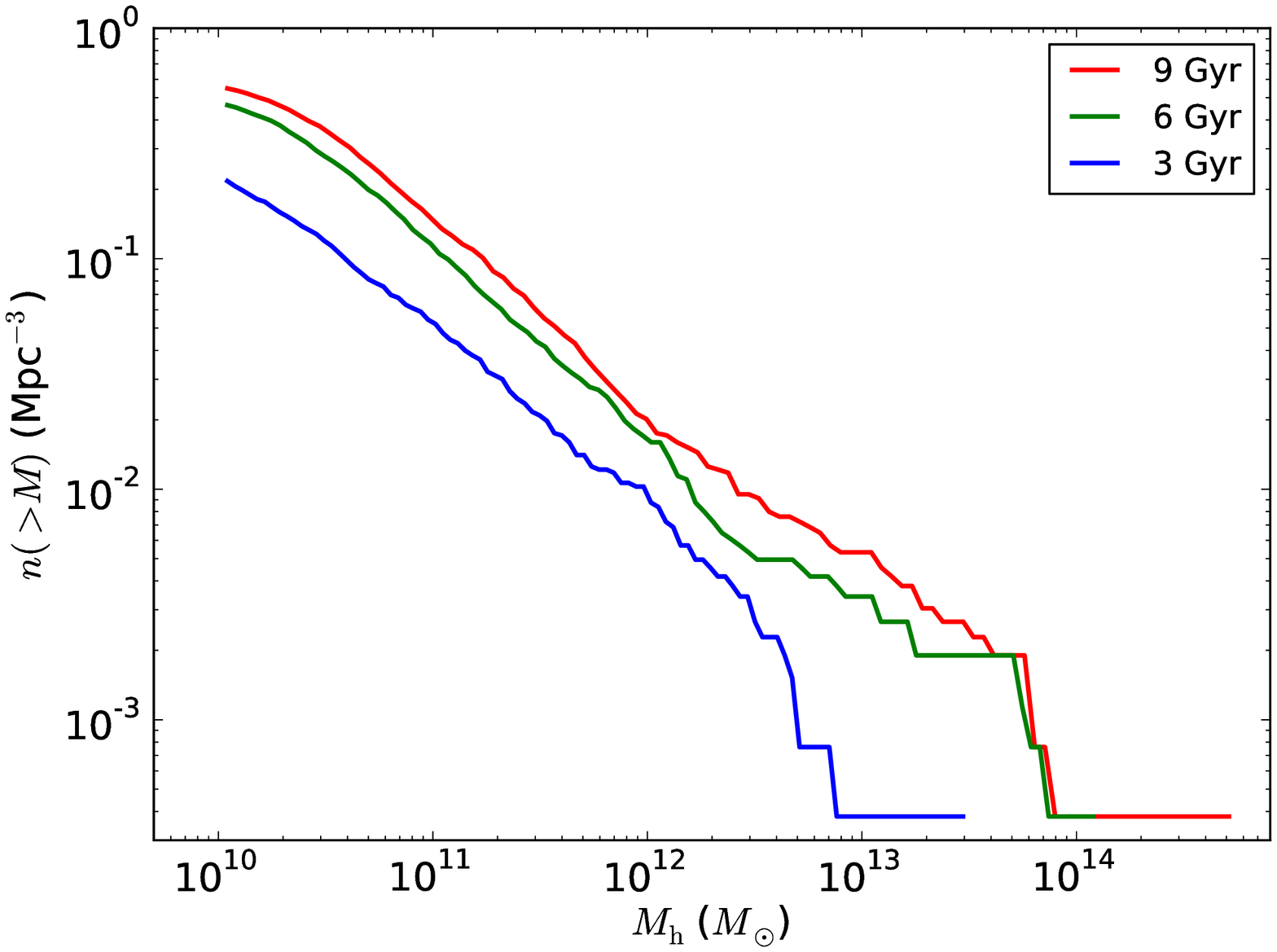}
      \label{fig:mspec_dm}
    }
    
    \caption{
      \label{fig:mspec}
      Cumulative mass  distribution of galaxies  and satellites (left)
      and haloes and subhaloes (right) at the third level of zoom,
      at $t=3$~Gyr (blue),  $t = 6$~Gyr (green), and  $t = 9$~Gyr (red)
      ($z = 2.29, 1.02,$ and 0.47).}
  \end{center}
\end{figure*}

Since  we built  the  merger trees  of  all galaxies  and dark  matter
haloes, we were able to  study the evolution of the mass function
with time. 
Figure~\ref{fig:mspec_bar}  and~\ref{fig:mspec_dm}  show,  respectively,
the  cumulative distribution  of mass  of  baryonic
galaxies and DM haloes at three timesteps of the simulation, the
three curves 
(blue, green, and red) corresponding to $t = 3, 6,$ and 9~Gyr,
respectively (or  $z = 2.23,  1.02,$ and 0.47),  in our third  level of
zoom.  
We computed the mass of structures and substructures therein, counting
the mass of substructures several times: once as stand-alone 
substructures, and then as part of their host structures.
The evolution of  the mass function is compatible  with a hierarchical
growth of structures, with  fewer massive structures, both in galaxies
and DM haloes,  existing at higher redshift than at  lower, and with a
slope that flattens towards lower redshifts. 
However, it  must be emphasised  here that the resulting  mass functions
are biased we ahve zoomed into an overdense region.
These      results     could      however      be     compared      to
\citet{2009MNRAS.399.1773C},  who performed  resimulations  of several
regions of the Millennium simulations.

\smallskip

\begin{center}
  \begin{figure}[t!]
    \includegraphics[width=8.5cm]{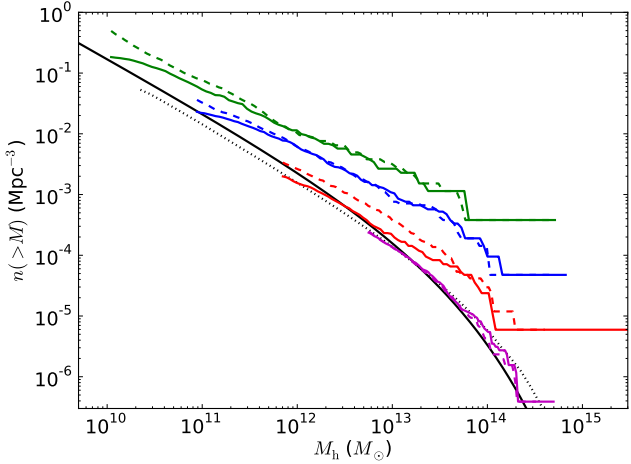}
    \caption{
      \label{fig:4lev} {AdaptaHOP (solid line) and FOF (dashed line)
      halo mass function for the
      four zoom level: level 3 in green,  2 in blue, 1 in red, and 0 in
      magenta. Note that subhaloes are not counted separately, but are
      included within  the AdaptaHOP haloes  in order to  compare with
      FOF haloes. 
      The black line is a  Press \& Schechter predicted mass function,
      for  the  sake  of   comparison,  and  characterisation  of  the
      environment.}  
    }
  \end{figure}
\end{center}

To  compare  several  structure-finding  codes, we  performed  another
structure detection with FOF, using  a linking length of $b=0.2$ times
the mean interparticular distance.  
Figures~\ref{fig:4lev} shows the  halo mass
function at the four zoom levels, respectively, for FOF (dashed line) and
AdaptaHOP (solid line) haloes. 
This time we note that the subhaloes are not counted separately, but are
included within the AdaptaHOP haloes to permit us to compare them with
FOF haloes.
Level 3 is shown in green, level  2 in blue, level 1 in red, and level
0 in magenta.
The black  solid line is a  Press and Schechter  function, computed by
the  code  described  in~\citet{2007MNRAS.374....2R} for  our  adopted
cosmology, and the  dotted line the mass function  from the Millennium
simulation \citep{2005Natur.435..629S}.  
The FOF  and AdaptaHOP mass functions show  differences, especially at
low masses. 
Indeed, several small haloes that are detected by FOF and located close to the
edges of a larger FOF halo are detected by AdaptaHOP as substructures
of this  halo, thus they do not  appear as low mass  structures in the
AdaptaHOP curve.  
However, at  higher masses there is  a good agreement  between the two
structure finders. 
Only  the level  0 (cosmological  run) mass  function can  be directly
compared with the 
theoretical predictions of Press \& Schechter and with the Millennium
mass function, although it is instructive to overplot the mass
functions for the three other simulations. 
The comparison with the Press  and Schechter mass function can be seen
as a probe of our environment and a way to quantify the overdensity.
Our mass function agrees with both the Millennium and  the Press
\& Schechter mass functions. 
The mass  functions in the other  zoom levels show the  density of the
environment with respect to the cosmic average. 

\subsection{Baryonic fraction}
\label{sec:barratio}

According to WMAP  data \citep{2011ApJS..192...18K}, baryons represent
about 4\% of the Universe content, yet only a small fraction are seen. 
One can legitimately ask where the baryons are in the Universe.
We computed the baryonic fraction, i.e. the ratio of the baryonic
mass to the DM mass, for each halo in the simulation, at several
timesteps. 
We defined  the halo centre as  the position of  the densest particle,
and the radius  of a (sub)halo as the distance  between the centre and
the furthest-away particle.  
For each halo or subhalo detected, we computed the total baryonic mass
within and plotted the baryonic fraction $m_\text{b}/m_\text{halo}$ as
a function  of the  (sub)halo (including all  its subhaloes)  mass, as
discussed in \S~\ref{sec:massfunc}.  
For each baryonic particle, we  computed  its closest dark matter
particle,  and assigned  the  baryonic  particle to  the  halo of  its
corresponding DM particle. 
This method enabled us to consider any geometry of halo. 
Haloes  undergoing major  mergers, which is  the case  for our
largest halo, may indeed have a non-spherical shape.

\smallskip

\begin{figure*}[t!]
  \begin{center}
    \includegraphics[width=17cm]{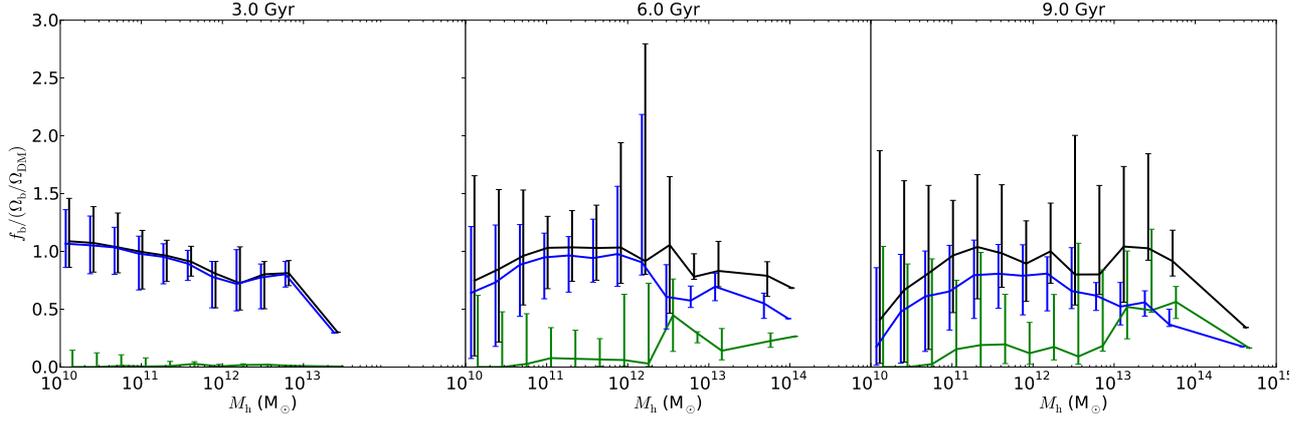}
    \caption{Median of  the baryonic fraction  computed in logarithmic
      bins as a function of the  halo mass in the level 3 zoom at
      $t=3, 6,$ and $9.0$~Gyr, normalised by the 
      universal fraction $\frac{\Omega_\text{b}}{\Omega_\text{DM}}$.
      The  total baryonic  fraction  is shown  in  black, the  stellar
      fraction in green, and the gas fraction in blue. 
      The errorbars represent the 85\nth\ and 15\nth\ percentiles.
      \label{fig:bfrac}}
  \end{center}
\end{figure*}
{

The baryon fraction, or the ratio of stellar mass to dark mass as a
function of time and galaxy mass can be compared with that  expected
based on analyses of observations with the halo abundance matching
technique (HAM).  
This tool  was developped by \citet{2004MNRAS.353..189V}  and has been
used by several groups.  
Assuming  that there  is a  tight correspondence  between  the stellar
masses of galaxies  and the masses of their  host haloes, and matching
their  number density or  abundance, this  technique allows  to deduce
average  relations linking  the baryon  and dark  matter  growths over
time, given  that the observed  galaxy stellar-mass function,  and its
variation with redshift, is satisfied as input.  
\citet{2009ApJ...696..620C} show for instance that the stellar mass growth
is essentially due to both accretion and star formation, while the merger
process has little influence and  most massive galaxies must form
their stars earlier than less massive ones (downsizing).  
The  stellar  mass  fraction  with  respect to  the  universal  baryon
fraction reaches a  maximum of 20\% in haloes that  have a virial mass
of a few  10$^{12}$ M$_\odot$ at $z=2$, and the  maximum is reached at
lower halo mass with time, down to a few $10^{12}\,\Msun$ at $z=0$. 
Figure~\ref{fig:bfrac} shows  the median of the  baryon fraction (in
black), stellar  fraction (green), and gas fraction  (blue) computed in
logarithmic mass bins, at $t = 3,
6, \text{and } 9$~Gyr, and such  an evolution can be seen, with a peak
in the stellar fraction at $\simeq 10^{12}\,\Msun$, which is compatible
with \citet{2009ApJ...696..620C}. 
These values are to some extent compatible with the relation found by
\citet{2010ApJ...717..379B},   who  consider   in   more  detail   the
uncertainties and scatter caused by various assumptions.
However, unlike these  authors, we still have a  higher stellar fraction
at higher than at lower mass. 
This  can be  atribudet to  our neglect  of feedback  from AGN  in our
simulations, which  is thought to  be responsible for  preventing star
formation in massive haloes. 
Observations of galaxies in groups and clusters \citep[e.g.][]{2005ApJ...635...73H,
  2010ApJ...719..119D} also show that there is a drop of the stellar fraction
towards high masses.  

Interestingly,  this plateau  at high  masses is  compatible  with the
simple prescription  in \citet{2012MNRAS.422.1714N},  for which the  stellar-to-halo mass
relation is modelled by a power law.
}

\begin{figure*}[ht]
   \begin{center}
     \includegraphics[width=17cm]{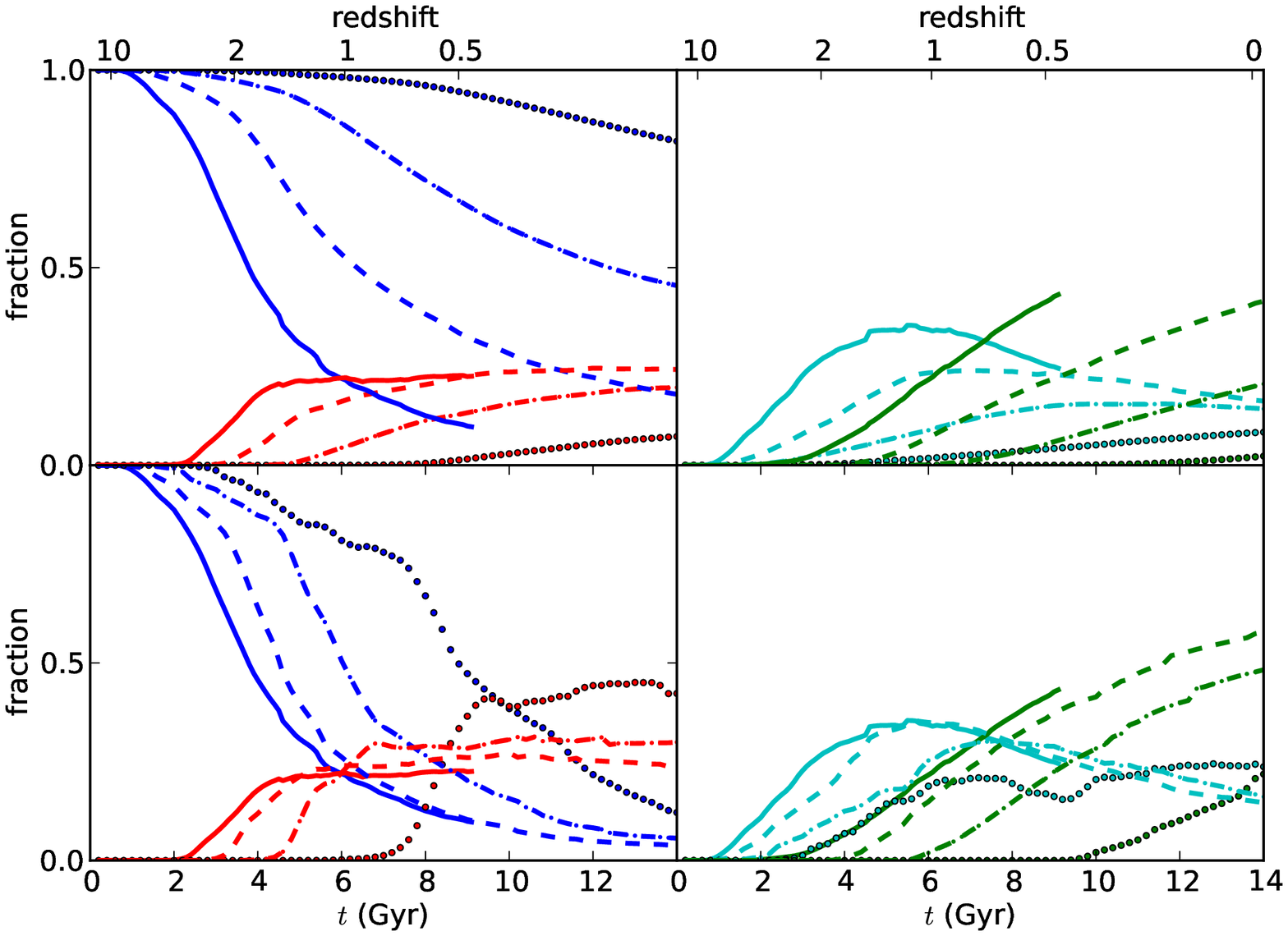}
     \caption{\label{fig:baryon_phase}Baryon  fraction in  four phases:
      diffuse  gas  (blue), condensed  gas  (cyan),  hot and  warm-hot
      (red)  and stars  (green), at  level 3  (solid), 2  (dashed), 1
      (dash-dotted), and 0 (dotted). 
      In the top panels, the fraction  is computed in the full box of
      each level, and in the bottom panels, it is computed only within
      the 8.56~Mpc radius of the level 3 box.  
      The differences are then only caused by those in the resolution.}
   \end{center}
 \end{figure*}
 
 \smallskip

We  studied the  relative  fraction of  baryons  in different  phases,
following~\citet{2001ApJ...552..473D}.
We distinguish  between four phases  according to the  gas temperature
and density contrast $\delta = \rho/\bar\rho - 1$.
\begin{itemize}
\item Diffuse gas: $\delta < 10^3$, $T < 10^5$~K.
\item Condensed gas: $\delta > 10^3$, $T < 10^5$~K.
\item Hot and warm-hot: $T > 10^5$~K.
\item Stars.
\end{itemize}

The evolution of each phase for the four levels of zoom is plotted in
Fig.~\ref{fig:baryon_phase}. 
Top panels show the  hot/warm-hot (red), diffuse (blue) gas, condensed
gas (cyan), and stars (green), computed for the four levels of zoom.  
These four  levels have a similar  trend of diffuse  gas that condenses
and forms stars  as cosmic structures evolve, and  the fraction of hot
and  warm-hot  gas  that  increases  when  there  are  massive  enough
structures to heat the gas, before eventually reaching a plateau.  
It is worth noting that the condensed gas fraction reaches a maximum
at  $z\simeq  2$,  which  corresponds   to  the  peak  of  the  cosmic
star-formation rate \citep[e.g.,][]{2010ApJ...709L.133B}.  

However,  the condensation  rate changes  from one  zoom level  to the
other. 
The difference  between the four levels  of zoom may be  due to either
the resolution  or the environment:  level 0 is indeed  a cosmological
box whereas  level 3 is  centred on a  dense region, and we  expect to
obtain different results owing to the different environments.
To distinguish both effects, we plotted in the bottom panels the same
fraction,  but  computed  for  each  level only  within  the  8.56~Mpc
spherical box of level 3.  
This time,  the difference between  the different zoom levels  are due
only to the resolution. 
We can see that there is a good convergence between levels 2 and 3.

\subsection{Dark  and orphan galaxies}

Since we  were able  to match galaxies  to haloes, we  checked whether
there was any ``dark  galaxy'', halo without any baryonic counterpart,
or ``orphan galaxy'', galaxy without a dark matter halo.

\smallskip

We  were unable  to  detect any  orphan  galaxy: all  galaxies in  the
simulation lie within a halo. 
However, not  all galaxies  are the  main galaxy of  either a  halo or
subhalo.  
We defined galaxies as the main structures identified with AdaptaHOP for
baryonic particles, and satellites to be their substructures. 
These  definitions differ  from those  usually assumed,  where central
galaxies  are at  the centre  of a  halo, and  all other  galaxies are
satellites.  
Therefore,   several detected structures     that  we   classified  as
``galaxies'' would be called ``satellites'' by other authors. 
Several of them are close to the halo centre, and cannot be associated
with a resolved subhalo. 
Whether this  is due to  a lack of  resolution or to  physical subhalo
stripping is still unclear.

\smallskip
 
We   detected   several  ``dark   galaxies''   without  any   baryonic
counterpart, about 100 haloes and 100 subhaloes at $t = 9$~Gyr.  
They are  represented in  figure~\ref{fig:bfrac} in cyan  (haloes) and
magenta (subhaloes).
We investigated whether {they contained any baryons} and found
that they appear to contain few gas particles.  
When looking at these dark  haloes and subhaloes, it appears that most
of  the  dark   haloes  have  gas,  but  that   these  structures  are
insufficiently well-resolved hence not  dense enough to form stars and
be detected as a galaxy.  
We therefore checked that these dark haloes and several subhaloes are
also  detected by FOF,  and are  not spurious  detections by  the halo
finder.  
Most of  the dark AdaptaHOP haloes  were also detected by  FOF, and one
can believe that they are of a physical significance. 
They  often contain  clouds  of gas  that  are not  dense  enough to  be
detected as  galaxies, which could  be an effect  caused by a  too low
resolution. 
{Some dark subhaloes were  also detected as FOF haloes, most
  of them  however appear to  be non-physical structures,  for example
  bridges between two real subhaloes containing a galaxy. }

\smallskip

By varying the sets of parameters in AdaptaHOP, we found different
numbers of dark galaxies. 
This  is  because  these  haloes  are  very  close  to  the  detection
threshold, and are detected as haloes  or subhaloes for a given set of
parameters, while they are undetected for a less conservative set. 

\subsection{Velocity dispersion}

\begin{figure*}[ht!]
  \begin{center}
    \subfigure[Galaxies]{
      \includegraphics[width=8.5cm]{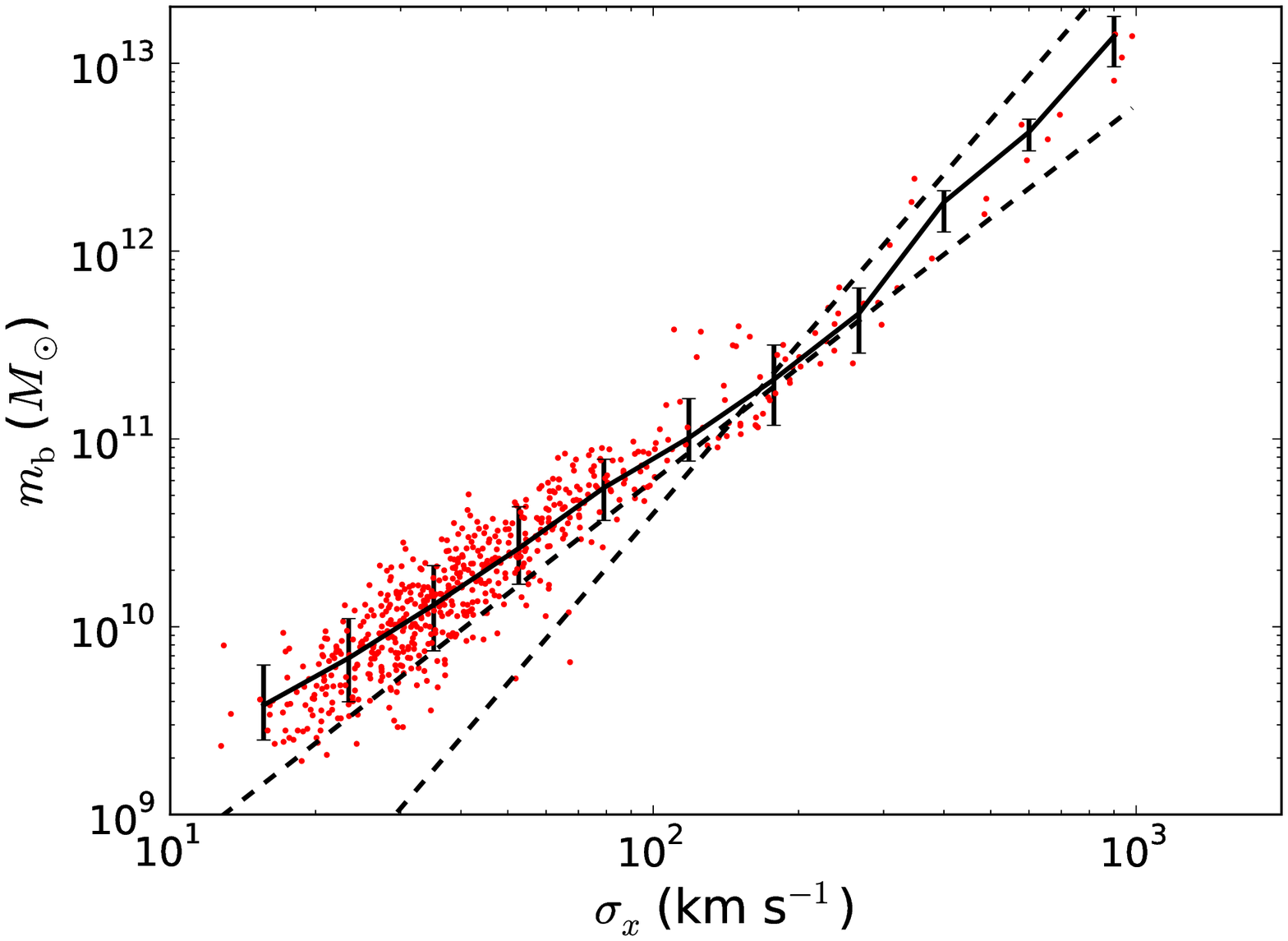}
      \label{fig:msigma_bar}
    }
    \subfigure[Haloes]{
      \includegraphics[width=8.5cm]{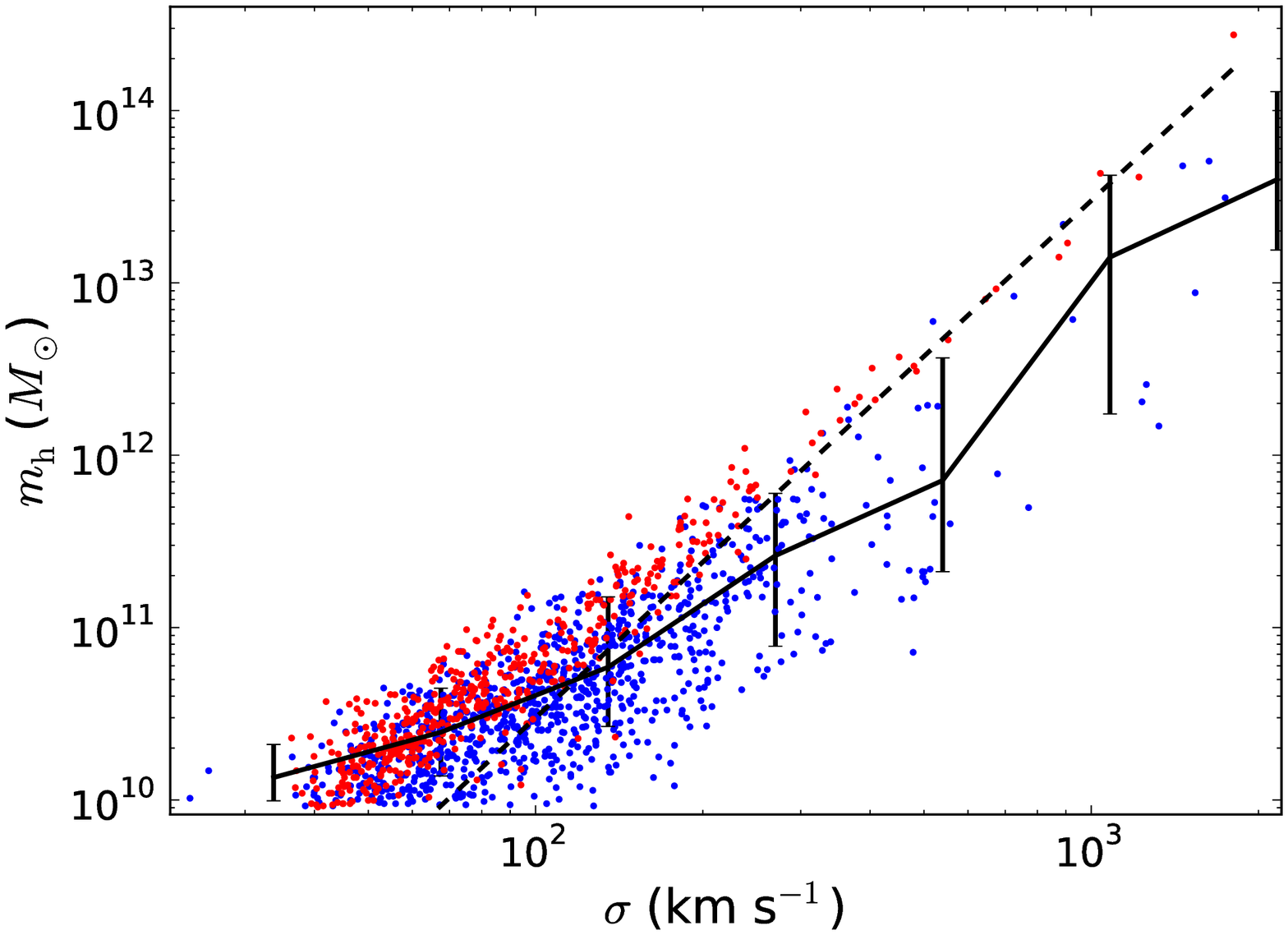}
      \label{fig:msigma_halo}
    }
    \caption{
      Mass versus velocity dispersion .  Left: galaxies; right:
      haloes(red) and subhaloes (blue). 
      The  solid black  line is  the median  of the  baryonic (stellar
      plus gas) mass of galaxies (panel
      \subref{fig:msigma_bar})  and  dark matter  mass  of haloes  and
      subhaloes (panel \subref{fig:msigma_halo}) and the errorbars are
      the 15\nth\ and 85\nth\ percentiles. 
      The dashed  black line  in the right  panel shows  the expected
      slope $m\propto \sigma^{3}$, and  on the right panel, the lines
      show the slopes $m\propto \sigma^{3}$ and $m\propto \sigma^{2}$.
      \label{fig:sigma_mass}
    }
  \end{center}
\end{figure*}

\begin{figure*}[ht!]
  \begin{center}
    \subfigure[Galaxy 2]{
      \includegraphics[width=0.45\linewidth]{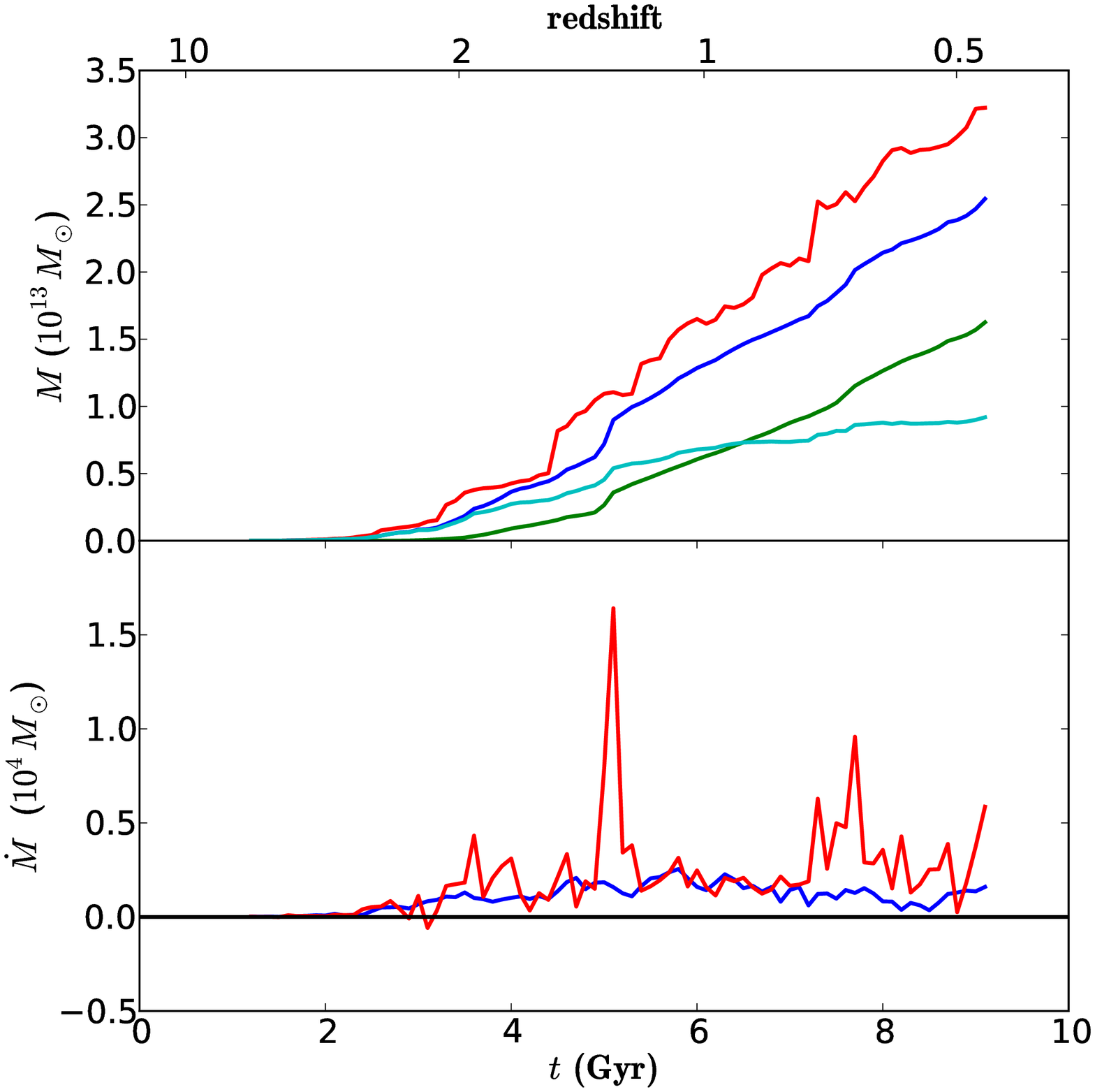}
      \label{fig:gal1}
    }
    \subfigure[Galaxy 56]{
      \includegraphics[width=0.45\linewidth]{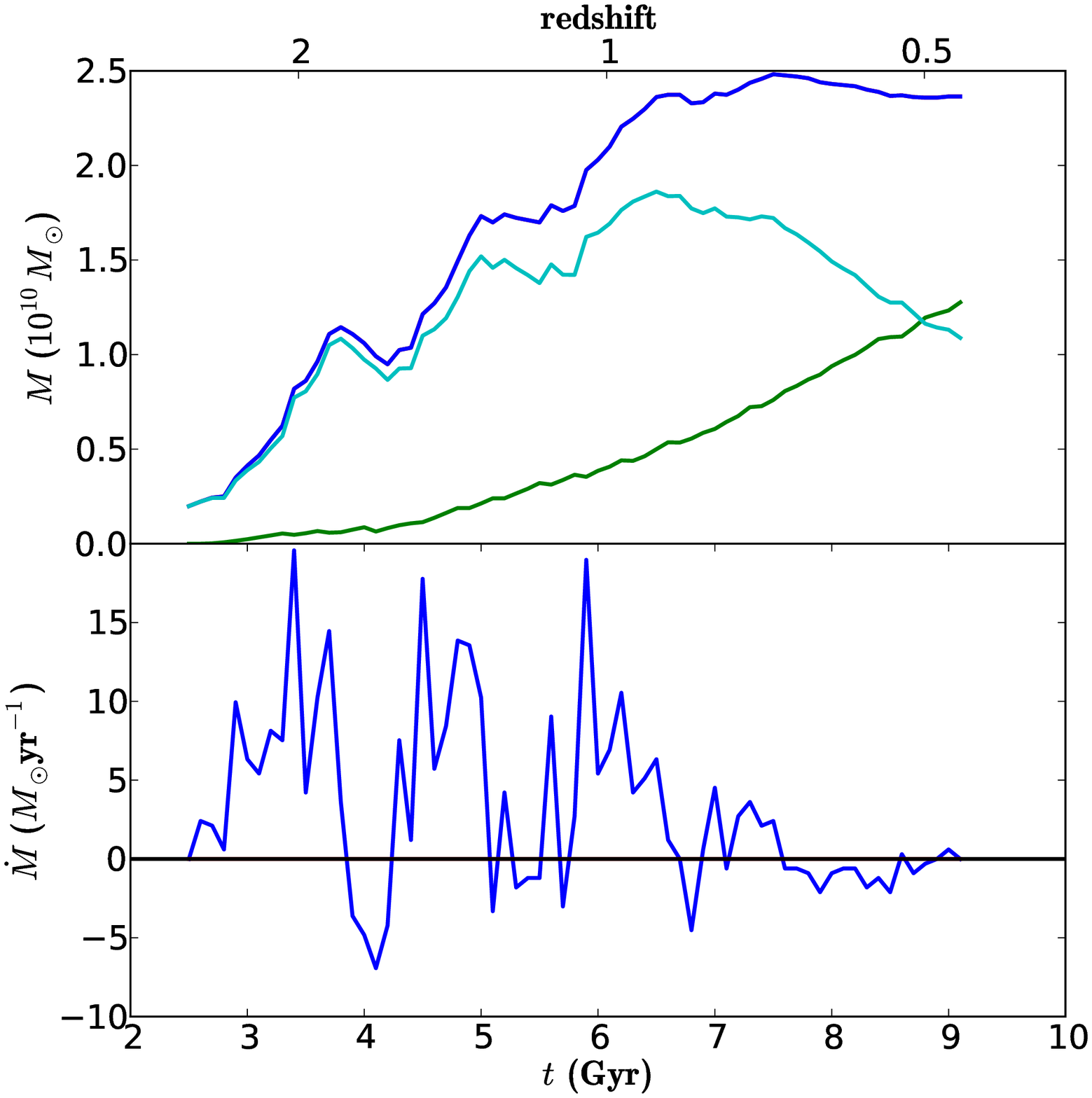}
      \label{fig:gal2}
    }
    \subfigure[Galaxy 469]{
      \includegraphics[width=0.45\linewidth]{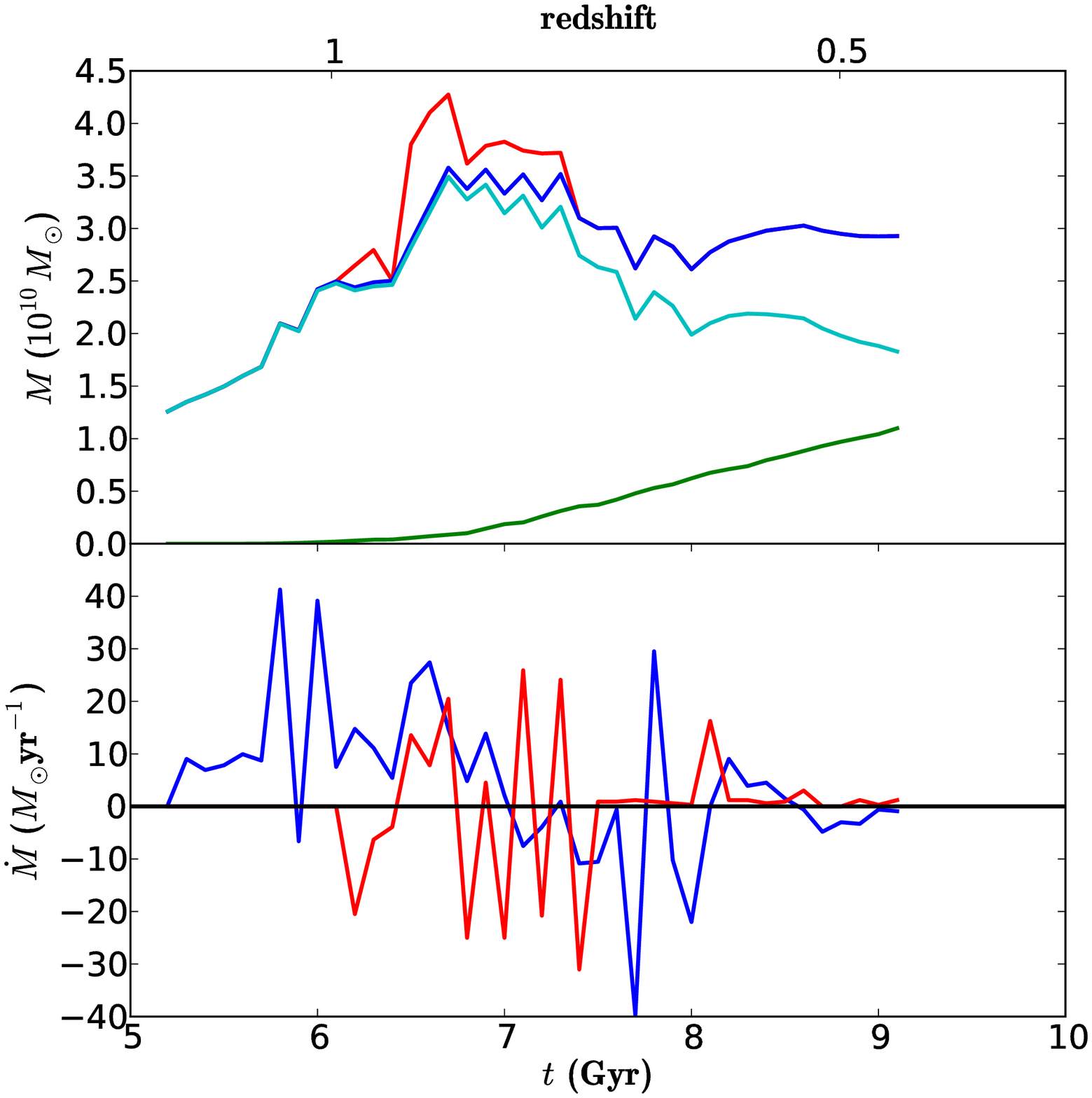}
      \label{fig:gal3}
    }
    \subfigure[Galaxy 512]{
      \includegraphics[width=0.45\linewidth]{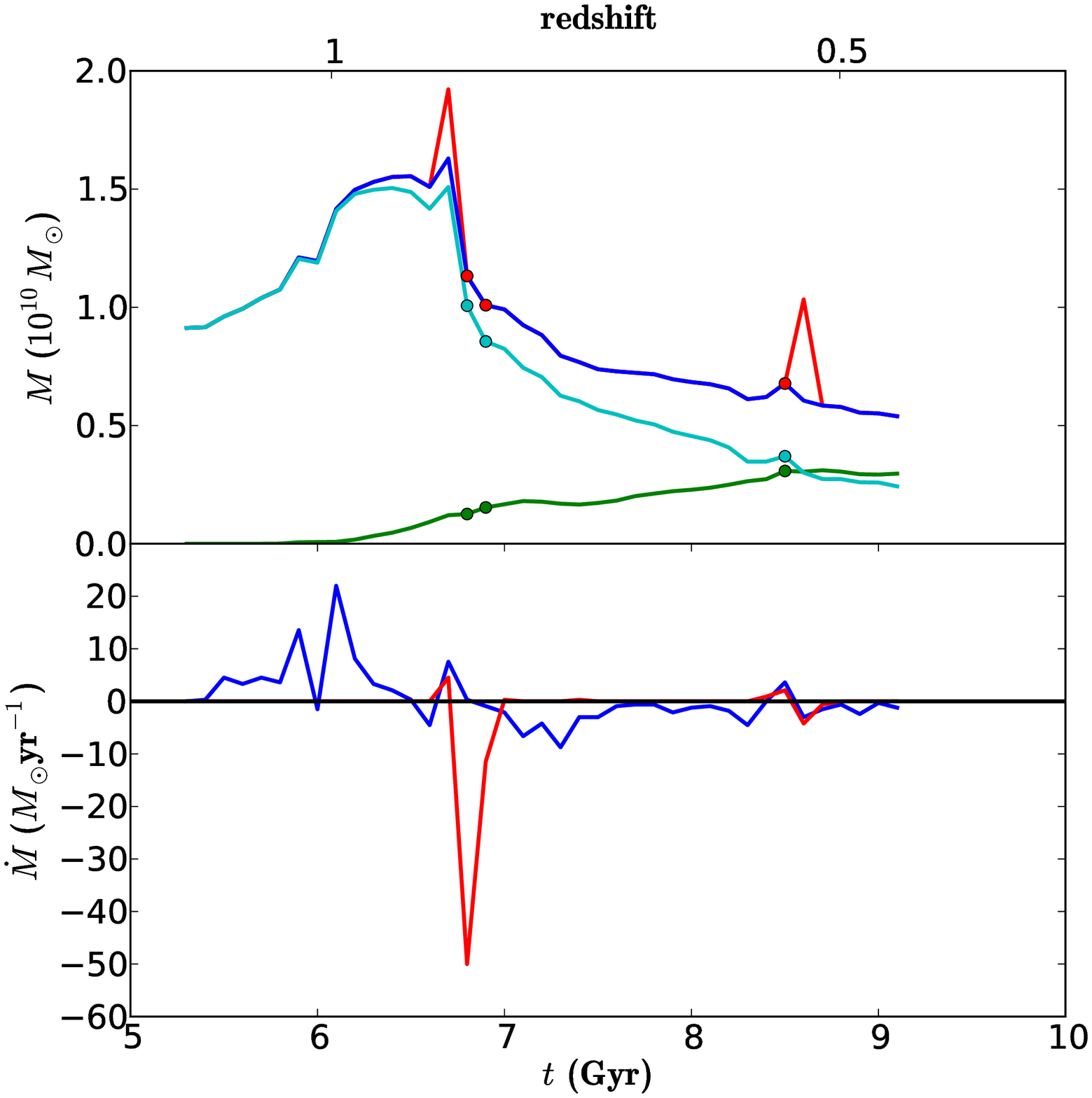}
      \label{fig:gal4}
    }
    \caption{ \emph{Top:}
      Mass history of four typical galaxies.  Blue curves: baryonic mass of
      the galaxy.   Red curves: baryonic  mass of the galaxy  plus its
      satellites. Green curve: stellar  mass of the main galaxy. Cyan:
      gas mass. 
      \emph{Bottom:} Mass  origin, where red represents  a merger from
      another  (sub)structure  and  blue,  smooth accretion  from  the
      background.  
      \label{fig:mass_hist}
    }  
    
  \end{center}
\end{figure*}

We computed  for each substructure, baryonic and  (sub)halo and galaxy
the velocity dispersion $\sigma_v$ at several snapshots. 
Since  AdaptaHOP does not  perform an  unbinding step,  some particles
that are spatially close to a structure can be attached to the latter,
although they are not dynamically bound.  
To get rid  of the contamination of these  particles, special care has
to be taken.
We used a two-step algorithm.
First, we  defined the  bulk velocity of  the structure by  computing the
median of  each component of the  velocity, which is  more robust than
taking the  mean value because  high velocity particles have  a weaker
influence on the median.
We then selected only particles with a velocity close enough to the
bulk velocity.
To do  so, we computed the  circular velocity at the  half-mass radius,
i.e. the radius containing half the mass of the structure,
$v^* =  \sqrt{\frac{GM(<r_\text{half})}{r_\text{half}}}$, which gives a
characteristic velocity for the structure.
{The  velocity dispersion  was then  computed  for particles
  whose velocity relative to $\vec{v}_\text{bulk}$ is lower than $n v^*$.
These are} the most bound particles.
We checked that  our result does not strongly depend  on the choice of
$n$, and took $n=5$. 

With this  technique, only one halo  and seven subhaloes  could not be
treated because no particle was selected. 
We checked  that these  structures corresponded to  unbound structures
and dropped them from our analysis.

{We  computed  the 3D  velocity  dispersion for  dark
  matter (sub)haloes.  
To  compare our  results  for baryonic  galaxies  and satellites  with
obsevations, we defined the velocity dispersion as follows: for massive
galaxies  ($M >  10^{11}$~\Msun), we  computed the  dispersion  in the
projected stellar velocity in radial bins along the $x$ axis, and defined
the  velocity dispersion  of the  structure as  the value  in  the bin
containing the half-mass radius.  
  For lower mass galaxies,  which are less well-resolved, the velocity
  dispersion is computed over all stellar particles. 
}

In this section, the mass of the structures does not take into account
the substructures, since we do not wish to account for the velocity
dispersion of particles within the substructures.

Figure~\ref{fig:sigma_mass} shows the  velocity dispersion as a function
of  the  mass for  galaxies  in  panel~\subref{fig:msigma_bar} and  DM
haloes (red) and subhaloes (blue) in panel~\subref{fig:msigma_halo}. 

{
      The dashed  black line  in the right  pannel shows  the expected
      slope $m\propto \sigma^{3}$, and  on the right pannel, the lines
      show the slopes $m\propto \sigma^{3}$ and $m\propto \sigma^{2}$.

The  relation between  mass and  velocity dispersion  for  dark matter
haloes  is compatible  with a  power law  with the  expected  slope of
three, although our data suggest a slightly shallower slope.
Observations indicate  that there is  a tight correlation  between the
total baryonic mass of galaxies and $V_\text{max}$, the maximum of the
velocity  curve, which  is referred  to as  the  baryonic Tully-Fisher
relation (BTRF).  
However     the    observed    exponent     is    close     to    four
\citep[eg][]{2000ApJ...533L..99M, 2007ApJ...671..203C, 
  2012AJ....143...40M}. 
In our case, the slope agrees with the BTRF at high masses, but at low
masses it is lower than expected. 
}

\section{Accretion and merger history}
\label{acc}

\begin{figure}[t!]
  \begin{center}
    \includegraphics[width=8.5cm]{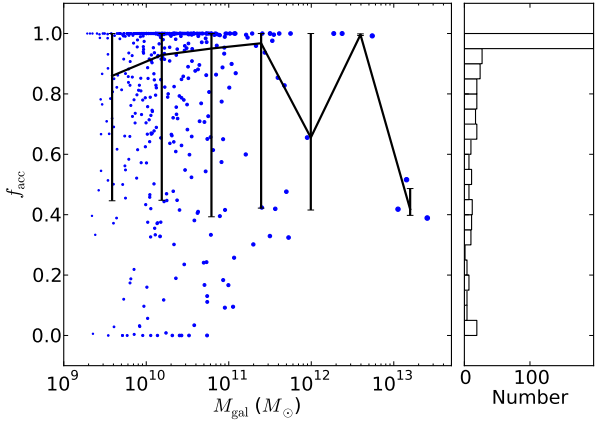}
    \caption{Accretion fraction versus galaxy mass, and the associated
      histogram (in number), computed for the 530 galaxies
      entering the box before 7~Gyr. 
      The size  of  the  markers  is proportional  to  the
      logarithm of the galaxy mass.
      In  the right  panel, red  dots correspond  to galaxies  that are
      outside the level 3 box.
      The black  line show the  median of the accretion  fraction, and
      the errorbars the 15\nth\ and 85\nth\ percentiles.
      We find a mean accretion fraction of 77\%.  
      \label{fig:hist_acc}}    
  \end{center}
\end{figure}

The upper panels of figure~\ref{fig:mass_hist} shows the mass history of
four characteristic galaxies of the simulation: a massive galaxy both
accreting gas  and growing  by mergers, a  small galaxy  growing only
through accretion, a galaxy growing by means of bowth mergers and
fragmentation, and a galaxy losing mass by means of fragmentation
while passing a larger galaxy.
The blue  curve is the baryonic  mass of the galaxy  itself, while the
red  curve shows  the  mass of  the  galaxy plus  its satellites.  The
stellar mass  is plotted  in green,  and the  gas mass  in cyan.
A galaxy may be detected as a satellite of another
galaxy during its history. These timesteps are plotted as circles
(panel \ref{fig:gal4}).

The bottom panels of figure~\ref{fig:mass_hist} show the origins of
the mass, separated into two components: merger and accretion. 
Smooth accretion is shown in blue and mergers in red.  
A negative  value of merger  or accretion, respectively, means  that the
galaxy loses mass to either another galaxy (fragmentation) or the
background (evaporation). 
These are the two components of the derivative of the blue curve shown
in the upper  panel, since with our definition,  all mass is acquired
by either merger or  smooth accretion, and lost by fragmentation or
evaporation.

Those four galaxies have very different mass accretion histories.  
The galaxy in panel~\subref{fig:gal1} undergoes a major 
merger that can  be seen in the lower  panel of \subref{fig:gal1}, at
$t \simeq 5.1$~Gyr.  

The galaxy in panel~\subref{fig:gal2} shows the opposite 
behaviour: 
it does not experience any merger and grows smoothly by accreting gas
until $t\simeq  7$~Gyr, then maintains a  constant mass until  the end of
the simulation, passively turning its gas reservoir into stars.
 
In panel~\subref{fig:gal3}, the galaxy grows mainly through accretion
until  it  reaches  a  maximum  mass at  $t\simeq  7$~Gyr,  and  then
interacts with another structure and 
loses more  mass than it gains  from mergers, and ends  with a somewhat
lower  mass. 

Galaxy  in panel~\subref{fig:gal4} shows  quite an  unusual behaviour:
after entering the level 3 box at $t \simeq 5\text{ Gyr}$, it grows
 from both mergers and accretion, and is suddenly accreted by a more
massive galaxy, becomes a satellite, then loses about one third of its
mass, which feeds the host galaxy. 
It  then leaves its  host galaxy  and continues  to lose  mass through
evaporation. 
These  behaviours  are quite  typical  of what  we  can  see in  our
simulations, with  high-mass  central galaxies undergoing  mergers and
accreting,  and lower-mass,  isolated galaxies  accreting gas  before
their growth is stopped. 
\smallskip

We computed the accretion fraction by considering, at the last output,
the origin of each particle:  particles belonging to the galaxy at the
first time of detection were defined as 
``initial''; particles that came  originally from the background, and
had never belonged to another structure were
labelled ``accretion'';  and we defined as ``merger''  a particle that
had  been  accreted  into  the  main  progenitor  of  the  galaxy  and
previously belonged to another structure, {even if it is coming from
the background.  }
We defined the  accretion fraction as
\begin{equation}
  f_\text{acc} = \frac{\text{accretion}}{\text{accretion + merger}},
\end{equation}

\noindent and the merger fraction $f_\text{merg}$ such that
$f_\text{acc}  + f_\text{merg}  =  1$. By  definition,  we have  $0\le
f_\text{acc},f_\text{merg}\le 1$. 

As a consequence  of the multi-zoom thechnique, and  that galaxies can
enter 
the level 3 box during the simulation, we had to select galaxies to be
studied. 
We only  computed this accretion  fraction for galaxies that  could be
tracked back  in time until  before $t =  7$~Gyr so that  the fraction
could be computed for at least 2~Gyr of its lifetime.  
We discarded  galaxies that could  not be tracked earlier  than 7~Gyr,
which  either
entered the box later, or were lost when computing the merger tree,
possibly after a merger event.
This  left   us  with  530   galaxies  that  had  been   tracked  from
$t_\text{app} < 7.0$~Gyr to $t = 9.1$~Gyr.

Figure~\ref{fig:hist_acc} shows the accretion fraction as a function of
mass for these 530 galaxies, as well as the (number) histogram of the accretion
fraction.  
We found a mean accretion fraction of 77\%, and a median value of 92\%.  
In black, we plotted the median accretion fraction in mass bins, where
the errorbars are the 15\nth\ and 85\nth\ percentiles.
We  can   see  that   most  galaxies  have   a  very   high  accretion
fraction. 
The trend  for low-mass galaxies  at $f_\text{acc}= 1$  indicates that
several galaxies undergo no mergers and are fed only by accretion. 
This could be a spurious effect caused by galaxies entering the box at
late time, and experiencing no mergers. 
However, even when we consider  only galaxies that are present in the
box between 3~Gyr and 9.1~Gyr, the histogram still shows such a trend with
a mean value of 70\% and median of 82\%.  
{ The  four points  in the upper  right region  are particularly
  striking:  they  correspond  to  massive galaxies  that  would  have
  acquired their mass mostly smoothly. 
When sudying the details, they correspond to galaxies that entered the
level 3 box a few snapshots before our limit of 7 Gyr, and  have
experienced   no  merger   since  this   date,  hence   have   a  high
accretionfraction.  
However, they are likely to have undergone mergers before entering the
level 3 box.
}

\section{Downsizing}
\label{sec:downsiz}

\begin{figure*}[ht!]
  \begin{center}
    \includegraphics[width=17cm]{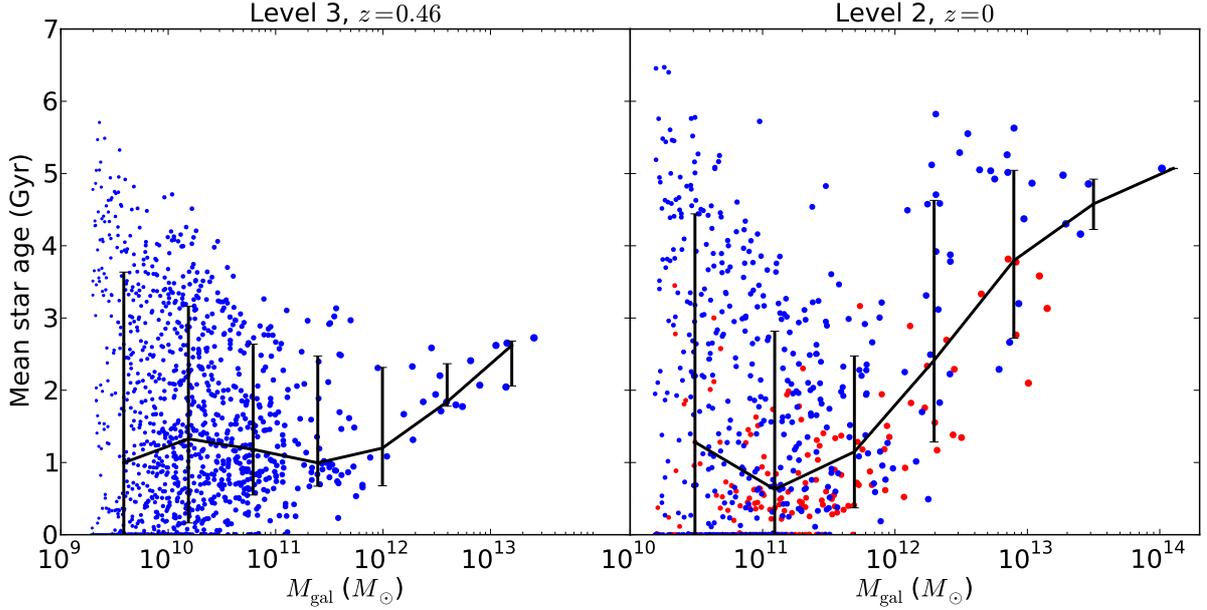}
    \caption{\label{fig:downsiz}Median of the mean stellar age of the galaxies as a
      function  of the  galaxy mass  at the  last output  for  level 3
      ($z=0.46$) and 2 ($z=0$).  
      The markersize is proportional to the logarithm of the mass.  
      In  the right  panel,  the blue  points  correspond to  galaxies
      within the $R_3 = 8.56$~Mpc of  the third level of zoom, and the
      red points to galaxies outside this region. 
      The errorbars show the 15\nth\ and 85\nth\ percentiles.
    }
  \end{center}
\end{figure*}  

We now  study the mean  stellar age of  galaxies as a function  of the
galaxy mass. 
For this study, we use both zoom levels two and three because the level
three simulation stops before $z=0$, but has a higher mass resolution.
Figure~\ref{fig:downsiz} shows  the median of the mean  stellar age of
the galaxies as  a function of the galaxy mass at  the last output for
level 3 ($z=0.46$) and 2 ($z=0$). 
The errorbars correspond to the 15\nth{} and 85\nth\ percentiles.

The differences between the blue dots  in the two panels are then due
bowth to our low resolution and  the temporal evolution. 
It is  interesting to see  that in the  second level of  zoom, galaxies
outside  the $R_3 = 8.56$~Mpc radius (i.e., red points) formed their
stars more recently than the ones within the box (blue points). 
This is  certainly due to the  lower average density of  the region in
the level 2 simulation which is located outside of the level 3 box.
We can see that at low  masses, the dispersion is large: among the least
massive galaxies, some form their  stars early, while others form them
late.  
At each epoch of the universe, there is a large number of dwarfs,
whose stars are forming actively. 
This behaviour is not seen for massive galaxies. 
The scatter  in the  stellar age progressively  reduces with  time, as
mass increases.  
The most  massive galaxies  form their stars  at a precise  epoch, 7-8
Gyr, which correspond to half  the universe age, depending slightly on
the level of resolution.  
 After this epoch, their  star formation drops considerably, which may
 be due to environmental effects that suppress the cold gas reservoirs.

This behaviour is compatible with both hierarchical structure formation,
since low-mass galaxies are the first  to form and then be involved in
the formation  of more massive galaxies, and  the observed downsizing,
since  the most massive  galaxies are  not observed  to form  stars at
$z=0$, but have formed most their  stars when the universe was half of
its present age.
Today, star formation continues only in small galaxies, although it was
also the case in the early universe.  

\section{Discussion}
\label{discuss}
\subsection{Influence of the resolution}
\label{sect:resol}
An  important question is  how variation  in the  numerical resolution
influences our results.
Our     study     of     the     baryon     phase     evolution     in
figure~\ref{fig:baryon_phase} provides  a first evidence  of numerical
convergence, especially between levels 2 and 3. 
To study the consistency in greater detail, we compared the mass assembly
history of several galaxies at zoom levels 2 and 3.
We identified these  galaxies in those two levels  of simulations, and
applying the same algorithms we compared their history.
Since the mass threshold is the  same, 64 particles, we expect that in
the  level 2 zoom,  fewer satellites  are detected  and sub-resolution
mergers play a role, thus the accretion fraction should be larger.

\begin{figure}[t!]
  \begin{center}
    \includegraphics[width=8.5cm]{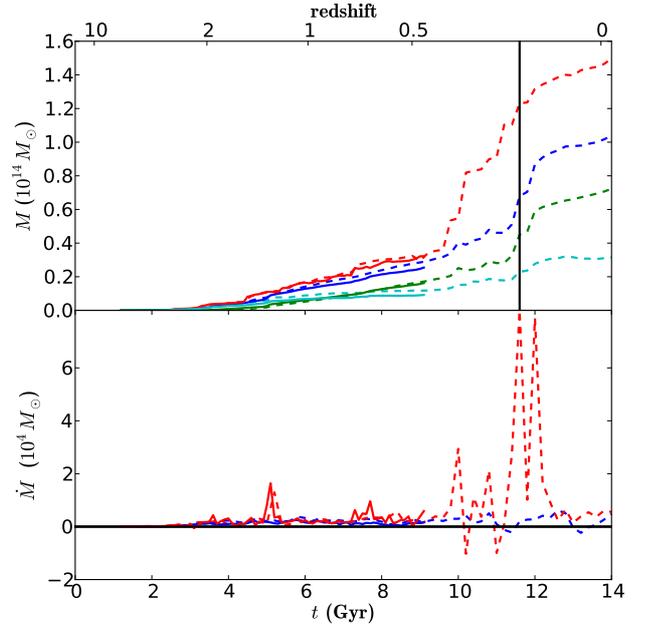}
    \caption{\label{fig:lev2} Mass assembly history of galaxy 1, same legend as
      figure~\ref{fig:mass_hist}. Solid line: level  3 zoom; dashed line: level
      2 zoom.}
  \end{center}  
\end{figure}

There are some galaxies for which the evolution is  far  from
complete at $t=9$~Gyr.  
We take the example of galaxy 1 in figure~\ref{fig:gal1}. 
Figure~\ref{fig:lev2} shows its mass assembly, the plain line is level
3 and the dashed line level 2. 
We can see  that the mass of  the galaxy is almost similar  in the two
zoom levels,  but the mass of the  satellites is lower in  the level 2
zoom. 

However,  in  the  level 2  zoom,  the  galaxy  is first  detected  at
$t=2.8$~Gyr,  whereas  in   the  level  3  one,  it   is  detected  at
$t=1.2$~Gyr, owing to a lack of resolution at level 2.

We found that the accretion fraction  between $ t= 0$ and $t = 9.0$~Gyr
is 0.64 at level 3 and 0.81 at level 2.
However, since level 2 reached $z=0$, we were able to compute the accretion
fraction for the entire formation history of the galaxy.
We found that $f_\text{acc} = 0.53$, which is lower than the accretion
fraction computed until $t=9.0$~Gyr. 
This  is  because this  galaxy  undergoes  major  mergers at  $t\simeq
12$~Gyr as we can see in figure~\ref{fig:lev2}.
We  also  computed  the  formation time  $t_\text{form}  =  11.6$~Gyr,
i.e. the time at  which the galaxy assembled half of its  mass at $z =
0$, which we show as a vertical black line in the figure. 
For  this galaxy,  owing to  the  late major  merger, $t_\text{form}$  is
greater than 9.1~Gyr, which means that it has assembled less than half of
its mass; however,  most galaxies have a formation  time that is lower
than 9.1 Gyr. 

{
\subsection{Overcooling issue}
\label{sec:agn}
As pointed  out in section~\ref{sec:ic}, no AGN  feedback was included
in this simulations serie. 
Active  galactic nucleus  feedback  has been  invoked  to resolve  the
over-cooling problem in massive galaxies. 
This could be done in several ways, and the complete issue has not yet
been definitely settled.  
At late times  (low redshift), when large structures  have formed, the
so-called ``radio mode'', coming from radio jets emitted by 
super-massive black-hole, is certainly efficient in preventing cooling
flows     \citep{2006MNRAS.365...11C,     2008MNRAS.390.1399B},    but
theirefficiency is believed to be local, and the 
action of AGN at higher redshift, (or the quasar mode) even before the
formation of groups  and clusters, is thought to  be more effective in
preventing the over-cooling \citep[e.g.,][]{2011MNRAS.412.1965M}.
Studies       taking       into       account       AGN       feedback
\citep[e.g.,  ][]{2011MNRAS.413..101G} obtain  more  realistic stellar
masses for massive haloes. 

Supernova feedback  is efficient for  low-mass haloes, where  kinetic energy
can be transferred into the gas, enabling it to escape the halo. 
However, the present simulation is focused on a massive cluster, where
the escape velocity is so high that SNe feedback is insufficient to
expel the gas from the halo. 
This results in the overcooling of baryons and eventually to very
high   stellar   masses  in   our   more   massive  haloes   (about
$10^{13}$~\Msun). 
We believe that, although  some results might suffer from overcooling,
our predictions should be robust at  least for the range of lower mass
galaxies. 
In a future paper, we will include various forms of AGN feedback and
address how our main conclusions should be modified as a consequence. 

}

\subsection{Comparison with previous work}

We found  that galactic  mass assembly is  dominated by  gas accretion
rather  than by  mergers, even  though we  might be  unable  to detect
mergers of low mass satellites.  
However,    we   do    not    expect   them    to   add  a  significant
contribution~(\citealt{2002ApJ...571....1M},
\citetalias{2005MNRAS.363....2K}).  
We note that we cover a comparable volume, with a comparable resolution
to \citetalias{2005MNRAS.363....2K},
although the main  advantage of our simulation is that  we are able to
simulate a rich environment starting from a cosmological box. 

Our      results       are      in      good       agreement      with
\citetalias{2005A&A...441...55S},   who  found   a   typical  accretion
fraction of 70\%.
We also appear to  agree with \citet{2009MNRAS.399..650S}, who studied
the mass growth of central and satellite galaxies.  
However, we  do not have the  same definition of  central galaxies and
satellites. 
They call a central galaxy the most  massive galaxy at the  centre of a
FOF-halo,  and all  other galaxies  lying in  the halo  or one  of its
subhaloes are called satellites.  
As a consequence, hey have  ``central'' galaxies of subhaloes that are
still accreting satellites, but  with their definitions these galaxies
are considered as satellites. 
\citet{2011MNRAS.tmp..662N}  studied the  satellite loss  mass in  a SPH
simulation  of a  Milky-Way  type galaxy  and  its satellites  through
UV-ionisation,  ram  pressure stripping,  stellar  feedback, and  tidal
stripping.  
They found that tidal stripping reduces significantly the
mass of bright satellites, which become of lower mass than some dark
ones,   and   that  stellar   feedback   mostly  affects   medium-mass
satellites. 
\Citet{2011MNRAS.tmp..554V}  performed  a   similar  study  using  the
Gadget-3 TreeSPH OWLS simulations. 
They computed the accretion rate onto both dark matter haloes and 
galaxies,  and  found  that  the  cold  accretion  dominates  the  mass
assembly. 
However, we note that we have not separated gas accretion into two different
components, namely  cold and hot  accretion, and considered  here only
accretion as opposed to mergers. 
\citet{2011MNRAS.417.2982F} also found that the contribution of mergers
to mass assembly is minor, and confined to high mass haloes.
They studied  in detail the  fate of accreted gas,  through cosmological
hydrodynamical simulations.  
They confirmed that  galaxies accrete mostly warm and  hot gas above a
critical halo mass of $3 \times 
10^{11}$~\Msun{} at $z=0$  as proposed by \citet{2003MNRAS.345..349B},
and that the fraction of cold gas accretion increases with redshift.  
Their variation in the  SNe feedback and efficiency of galactic
winds demonstrated  that this essentially  unknown parameter can  decouple the
gas accretion  from the star formation rate,  and efficiently decrease
the baryon fraction in low-mass haloes.

 \smallskip

Interestingly,   \citet{2010ApJ...725.2312O}  studied
stellar assembly,  distinguishing between ``in situ''  stars that were
formed within  the galaxy, and ``ex  situ'' stars that  were formed in
another galaxy before entering the current one.  
Although these definitions differ somewhat from ours, they are closely
related, since ``in situ stars'' are locally formed from cold gas, and
``ex situ'' are assembled by mergers.
They found that ``in situ'' stars dominate  low mass galaxies  at
earlier  times, while  massive galaxies  are dominated  by  ``ex situ''
star accretion, and  in their case the formation  of ``in situ'' stars
occurs as the result of the accretion of cold flows.
We agree--at least  qualitatively--with their evidence of downsizing
and  observe a  similar trend  in mean  stellar age  as a  function of
galaxy mass,  although they have a smaller scatter in the  mean stellar
age.  

\citet{2011A&A...533A...5C}   performed   a   similar   analysis   using
semi-analytical methods. 
They found that galaxies  less massive than $10^{11}\,\Msun$ assembled
most of their mass by mergers rather than by accretion.
We qualitatively agree with  their results, even though our statistics
for massive galaxies  are of insufficiently high quality  to draw firm
conclusions. 

Using high-resolution dark  matter simulations (the Aquarius project),
\citet{2011MNRAS.413.1373W} also found that mergers with mass ratios
larger than  1:10 contribute  very little to  the DM  mass assembly,
less than 20\%.  
Most  of the  major merger  contribution  is confined  to the  central
parts of haloes, which does not represent the bulk of the mass. 
This investigation can be extended to baryons, through semi-analytical
prescriptions                       \citep[e.g.][]{2000MNRAS.319..168C,
  2006MNRAS.370..645B},  since  the lowest  mass  halos  have a  small
baryon fraction.

\smallskip

\section{Conclusions} 
\label{conclu}

We have studied the accretion histories of 530 galaxies using multi-zoom
simulations, starting from  a cosmological simulation and resimulating
three times smaller zones of interest at higher resolution. 
We selected  a dense region  and detected {the  hierarchy of
  dark matter haloes and 
subhaloes,  as well as baryonic galaxies  and satellites}, at
each  timestep of  the simulations,  which  enabled us  to follow  the
structures in time, and build merger trees. 

We computed the mass assembled through both smooth gas accretion and
mergers, and we found that accretion plays a dominant role, at
least  until  $z\simeq  0.4$,  the  end of  our  highest  zoom  level
simulation.  
Massive galaxies have a lower mass-accretion fraction. 
Over all galaxies, about three-quarters of the mass is on average assembled
through smooth accretion, and one-quarter through mergers.  
This is in agreement with previous studies examining the role of gas
accretion    \citep[eg,][]{2002ApJ...571....1M,   2003MNRAS.345..349B,
  2005A&A...441...55S,2005MNRAS.363....2K,         2009MNRAS.395..160K,
  2009ApJ...694..396B,  2011MNRAS.417.2982F, 2011MNRAS.tmp..554V}, but
we  have extended these  previous analyses  to achieve  higher quality
statistics. 

The  main originality  of  this work  lies  in the  use of  multi-zoom
simulations, which allow to simulate galaxies with a fairly high resolution
in a cosmological context, and to study their mass assembly history. 
It  is worth  noting that,  even  in quite  dense environments,  where
mergers  are expected  to occur,  mass assembly  is still  dominated by
smooth accretion.

\smallskip

We have also studied the evolution of the mass functions of galaxies and DM
haloes, especially in  dense environments.  
The galaxy  density in our  final zoom level  is in-between  group and
cluster environment, and the most massive galaxies are spiral (and not
elliptical) at the end of the simulation ($z=0$ at the level 2).  
This  evolution,  especially for  galaxies,  clearly  agrees with  the
hierarchical model,  with low-mass  galaxies at high  redshift and
massive ones at low redshift. 

\smallskip

We have been able to match DM haloes and galaxies, finding no
galaxies without haloe. 
In contrast, we did find some dark structures containing no galaxies,
although  this  is likely  to  be  an artefact  caused  by  a lack  of
numerical  resolution, and  with  a higher  resolution these  galaxies
would be detected.  
We have studied the baryonic content of haloes and found that lower mass
haloes have smaller  baryon fractions, as expected from  the action of
stellar feedback.  
{We have studied the evolution of the baryon phases in our simulations,
into  different   components,  namely  diffuse   gas,  condensed  gas,
hot/warm-hot gas, and stars, and distinguishing between the effects of
environment and resolution. }

\smallskip

Finally, we have studied the mean stellar age of our galaxies, at both
$z=0.47$ and $z = 0$, and  found evidence of downsizing: low mass
galaxies form stars  at each epoch, whereas in  massive galaxies, most
stars have formed when the universe was half its present age.

\smallskip

These results however  suffer from the problem of  overcooling that is
encountered at high masses. 
The  downsizing trend  should remain  detectable despite  this problem
although the accretion  fraction and the gas content  of haloes depend
on the adopted feedback recipes.
The  results  for the  more  massive  galaxies  should thus  be  taken
with caution.

\smallskip

We propose to develop further simulations to explore different physical
parameters such as feedback or star formation recipes in order to help
us  understand  their  influence  on  gas accretion  as  advocated  by
\citet{2011MNRAS.tmp..554V} and \citet{2011MNRAS.417.2982F}.
 In this  next series of simulations,  we will be able  to address the
 role of feedback at various epochs in determining the accretion fraction by
 comparing these simulations with SNe feedback only with our present results.

\smallskip

We note that  our present study has been performed  on a dense region,
and new multi-zoom  simulations centred on less dense  regions will be
analysed to help us ascertain the role of the environment.
The relative importance of mergers and diffuse accretion might indeed 
depend on environment, as \citet{2007ApJ...654...53M} have shown with
dark-matter only simulations.  
The time scale for the assembly of massive haloes is shorter in dense
environments, and the relative role of mergers appears to be higher. 
Environments with a wide range of densities must therefore be explored
by he use  of full simulations with baryons and  feedback to perform a
census of  mass assembly encompassing wide ranges  of environments and
redshifts. 
The geometry of  gas accretion from filaments onto  discs will also be
studied in more detail. 

\begin{acknowledgements}
We thank Dylan Tweed for providing us with tools to build merger trees and
help  in the use  of both  AdaptaHOP and  the merger  trees St\'ephane
Colombi  for  stimulating  discussions  on  structure  detection,  and
Ana\"elle Hall\'e for her comments on the paper.  
The simulations  were performed at  the CNRS supercomputing  center at
IDRIS, Orsay, France. 
\end{acknowledgements}

\bibliographystyle{aa} 
\bibliography{accretion}

\end{document}